\documentclass[aps,prd,twocolumn,amsmath,superscriptaddress,amssymb,showpacs,floatfix,nofootinbib,longbibliography]{revtex4-1}
\usepackage{graphicx}
\usepackage{dcolumn}
\usepackage{xcolor}
\usepackage{bm}
\usepackage{natbib}
\usepackage{calligra}
\usepackage[T1]{fontenc}
\usepackage{egothic}
\usepackage[T1]{fontenc}
\newfont{\rsfsten}{rsfs10 scaled 1200}
\newfont{\rsfsseven}{rsfs10 scaled 1200}
\newfont{\rsfsfive}{rsfs10 scaled 1200}
\usepackage{epsfig}
\usepackage{units}
\usepackage[utf8]{inputenc}
\usepackage{hyperref}

\newcommand{\aap}{{Astron.~Astrophys.}}
\newcommand{\apjl}{{Astrophys.~J.~Lett.}}

\newcommand{\mnras}{{Mon.~Not.~R.~Astron.~Soc.}}

\newcommand{\be}{\begin{equation}}
\newcommand{\ee}{\end{equation}}
\newcommand{\bea}{\begin{eqnarray}}
\newcommand{\eea}{\end{eqnarray}}

\def\lsim{\mathrel{\raise.3ex\hbox{$<$\kern-.75em\lower1ex\hbox{$\sim$}}}}
\def\gsim{\mathrel{\raise.3ex\hbox{$>$\kern-.75em\lower1ex\hbox{$\sim$}}}}

\begin{document}

\title{Observing Signals of Spectral Features in the Cosmic-Ray Positrons and Electrons from Milky Way Pulsars}

\author{Ilias Cholis}
\email{cholis@oakland.edu, ORCID: orcid.org/0000-0002-3805-6478}
\affiliation{Department of Physics, Oakland University, Rochester, Michigan, 48309, USA}
\author{Thressay Hoover}
\email{hoover92@purdue.edu}
\affiliation{Department of Physics, Oakland University, Rochester, Michigan, 48309, USA}
\affiliation{Department of Physics and Astronomy, Purdue University, West Lafayette, Indiana 47907, USA}
\date{\today}

\begin{abstract}
The Alpha Magnetic Spectrometer (\textit{AMS-02}) has provided unprecedented precision measurements 
of the electron and positron cosmic-ray fluxes and the positron fraction spectrum. At the higher energies, sources as 
energetic local pulsars, may contribute to both cosmic-ray species. The discreteness of the source population, can result in 
features both on the positron fraction measurement and in the respective electron and positron spectra. For the latter, those 
would coincide in energy and would contrast predictions of smooth spectra as from particle dark matter. In this work, using
a library of pulsar population models for the local part of the Milky Way, we perform a power-spectrum analysis on the 
cosmic-ray positron fraction. We also develop a technique to cross-correlate the electron and positron fluxes. We show that
both such analyses, can be used to search statistically for the presence of spectral wiggles in the cosmic-ray data. For a 
significant fraction of our pulsar simulations, those techniques are already sensitive enough to give a signal for the presence
of those features above the regular noise, with forthcoming observations making them even more sensitive. Finally,  
by cross-correlating the \textit{AMS-02} electron and positron spectra, we find an intriguing first hint for a positive correlation
between them, of the kind expected by a population of local pulsars. 

\end{abstract}

\maketitle

\section{Introduction}
\label{sec:introduction}
\vspace{-0.3cm}
In recent years, the fluxes of cosmic-ray electrons and positrons at GeV to TeV energies have 
been measured with an unprecedented accuracy by the Alpha Magnetic Spectrometer (\textit{AMS-02}), 
aboard the International Space Station, \cite{AMS:2019iwo, AMS:2019rhg}. These cosmic rays originate
from a sequence of sources and mechanisms. Most electrons get accelerated to these energies in supernova 
remnant (SNR) environments. As the SNR shock front expands outwards, particles in the surrounding 
interstellar medium (ISM), experience 1st order Fermi acceleration  \cite{PhysRev.75.1169, 1954ApJ...119....1F, 
Caprioli:2013dca}. These cosmic rays are conventionally referred to as primary cosmic rays. Primary cosmic-ray
electrons are only one of the particle species accelerated in SNRs. Other species include protons and more massive 
nuclei. These nuclei, as they propagate through the ISM, may have hard inelastic collisions leading to the 
production of charged mesons that once decaying will produce among other particles the secondary cosmic-ray
electrons and positrons \cite{Moskalenko:2001ya, Kachelriess:2015wpa,  Strong:2015zva, Evoli:2008dv}. The 
secondary electrons are an important component especially at the lower \textit{AMS-02} energies, while the 
secondary positrons are the prominent mechanism by which cosmic-ray positrons are produced in the Milky Way. 
However, multiple measurements including those from the \textit{AMS-02}, the Payload for Antimatter Matter 
Exploration and Light-nuclei Astrophysics (\textit{PAMELA}), the \textit{Fermi}-Large Area Telescope, the 
CALorimetric Electron Telescope (\textit{CALET}) and the Dark Matter Particle Explorer (\textit{DAMPE}), 
suggest the presence of an additional source of high energy electrons and positrons 
\cite{AMS:2019rhg, AMS:2019iwo, AMS:2021nhj, Adriani:2018ktz, DAMPE:2017fbg, PAMELA:2013vxg, 
Fermi-LAT:2011baq}. That is most notable in the positron fraction, i.e. the ratio of cosmic-ray 
positrons ($e^{+}$) to electrons ($e^{-}$) plus positrons ($e^{+}/(e^{+}+e^{-})$). That spectrum rises from
5 GeV and at up to $\simeq 500$ GeV in energy \cite{PAMELA:2008gwm, PAMELA:2013vxg, 
Fermi-LAT:2011baq, AMS:2014bun, AMS:2019iwo, AMS:2021nhj}.

The source of these high energy electrons and positrons has been debated since the first robust detection of
additional positrons by \textit{PAMELA} \cite{PAMELA:2008gwm}. One mechanism is that SNR environments can 
source also secondary cosmic rays that remain within the SNR volume for enough time to get accelerated 
before escaping into the ISM \cite{Blasi:2009hv, Mertsch:2009ph, Ahlers:2009ae, Blasi:2009bd, Kawanaka:2010uj, 
Fujita:2009wk, Mertsch:2014poa, DiMauro:2014iia, Kohri:2015mga, Mertsch:2018bqd}. However, such a mechanism
would also produce other species of high-energy cosmic rays that we have not observed at the expected fluxes
associated to the cosmic-ray positrons \cite{Cholis:2013lwa, Mertsch:2014poa, Cholis:2017qlb, Tomassetti:2017izg}.
Another type of cosmic-ray positrons sources is local pulsars  \cite{1987ICRC....2...92H, 1995PhRvD..52.3265A, 
1995A&A...294L..41A, Hooper:2008kg, Yuksel:2008rf, Profumo:2008ms, Malyshev:2009tw, Kawanaka:2009dk, 
Grasso:2009ma, 2010MNRAS.406L..25H, Linden:2013mqa, Cholis:2013psa, Yuan:2013eja, Yin:2013vaa, 
Cholis:2018izy, Evoli:2020szd, Manconi:2021xom, Orusa:2021tts, Cholis:2021kqk, Bitter:2022uqj}, while a third 
more exciting possibility is that of dark matter annihilation or decay in the Milky Way \cite{Bergstrom:2008gr, 
Cirelli:2008jk, Cholis:2008hb, Cirelli:2008pk, Nelson:2008hj, ArkaniHamed:2008qn, Cholis:2008qq, Cholis:2008wq, 
Harnik:2008uu, Fox:2008kb, Pospelov:2008jd, MarchRussell:2008tu, Chang:2011xn, Cholis:2013psa, Dienes:2013xff, 
Kopp:2013eka, Dev:2013hka, Klasen:2015uma, Yuan:2018rys, Sun:2020dla}.

SNRs come with a distinct age while pulsars are known to lose most of their initial rotational energy very fast. This makes 
both SNRs and pulsars with their surrounding pulsar wind nebula (PWN), sources that will inject into the ISM most of the high-energy 
cosmic-ray electrons and positrons on a short timescale after the initial supernova explosion. That timescale is of 
$O(10)$ kyr. This is a couple of orders of magnitude smaller than the time required for cosmic rays to reach 
our detectors at Earth. Thus, we expect that the observed high-energy spectra from such sources will display an energy 
cut-off due to cooling associated with their age \cite{Malyshev:2009tw, Profumo:2008ms, Grasso:2009ma, Cholis:2017ccs}. 
Unless only one close-by powerful source dominates the entire energy range observed by \textit{AMS-02}, different 
members of a population of sources should contribute at different energies. As electron and positron cosmic rays lose their  
energy fast at the highest energies, only a small number of astrophysical sources can contribute. If either a population 
of Milky Way pulsars or SNRs is responsible for the observed high-energy positrons, then we expect some small in 
amplitude spectral features to arise on the positron fraction above a certain energy. In fact, the positron fraction spectrum
cutting off at $\sim500$ GeV, may be suggestive of either running out of possible close-by pulsar/SNR sources or 
the properties of a dark matter particle. However, a dark matter particle (or even a more complex dark sector) would 
not predict any smaller features on the cosmic-ray positron fluxes \cite{Cholis:2009va, Dienes:2013xff}. In this work, we 
make use of that discriminant characteristic between a population of conventional astrophysical sources and dark matter. 
If a signal of small-scale spectral features is robustly detected at the higher end of the cosmic-ray positron flux and 
possibly electron flux observations, that would provide strong evidence against the dark matter interpretation. 

\begin{figure}
\begin{center}
\hspace{-0.5cm}
\includegraphics[width=3.7in,angle=0]{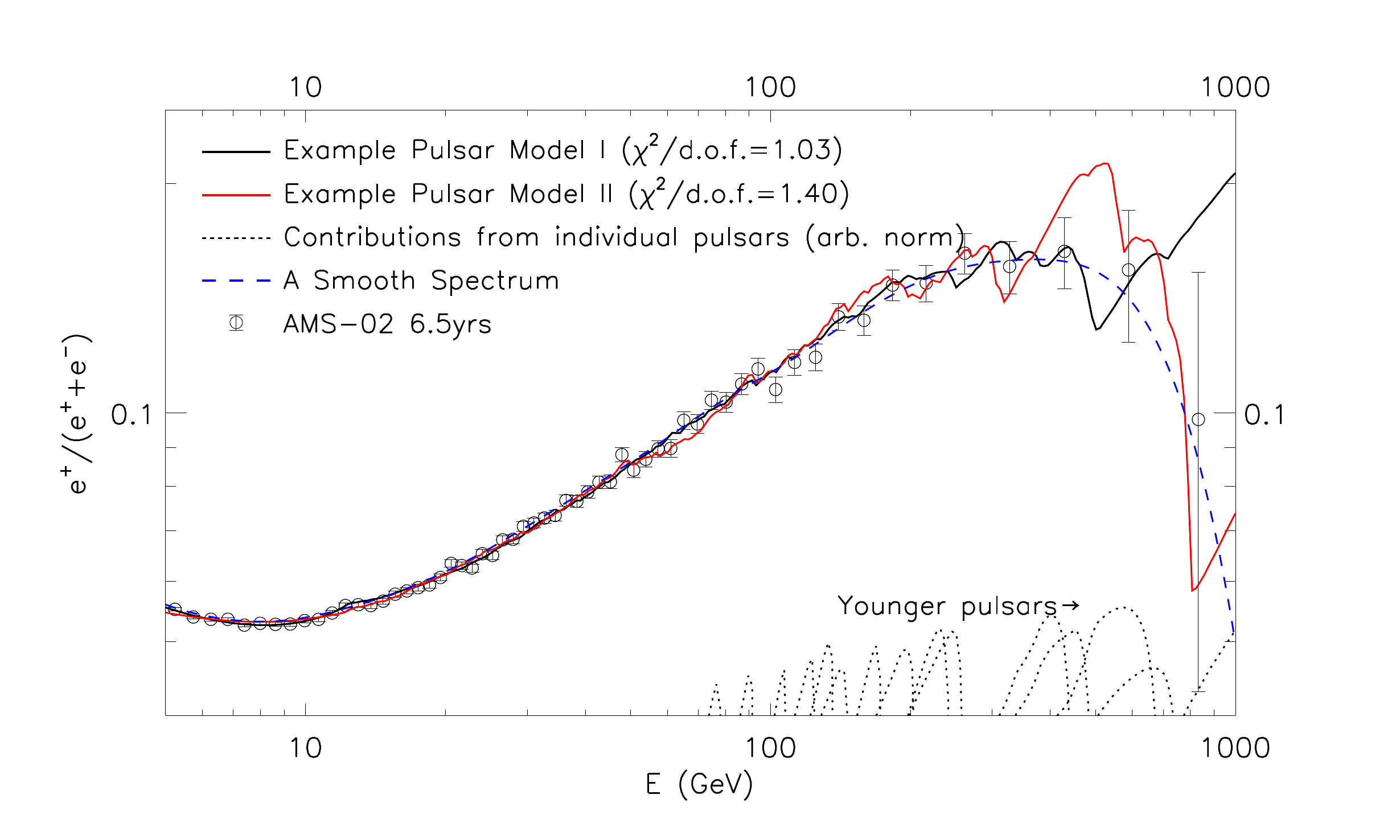}
\end{center}
\vspace{-0.7cm}
\caption{The \textit{AMS-02} positron fraction measurement (data points) of Ref.~ \cite{AMS:2019iwo}. 
We show in the solid black and red lines the positron fraction from two pulsar population models, originally produced 
in~\cite{Cholis:2021kqk} and in the dashed blue line a smooth positron fraction as that coming from an
annihilating dark matter particle. The contribution of individual pulsars is shown by the dotted lines. Younger 
pulsars contribute at higher energies. The flux normalizations from individual pulsars are taken to be arbitrary, just to 
showcase the contributions of some pulsars and are picked to be higher for the older pulsars in order
for them to appear within plot (see text for more details).}
\label{fig:PositronFraction}
\end{figure}

In Figure~\ref{fig:PositronFraction}, we show the \textit{AMS-02} positron fraction measurement of \cite{AMS:2019iwo}, 
together with sample pulsar models that can explain it, taken from Ref.~\cite{Cholis:2021kqk}. We also show a smooth 
positron fraction as that coming from an annihilating dark matter particle. Milky Way pulsar population models, typically 
predict the existence of small spectral features originating from the contribution of individual pulsars. The younger and 
more local pulsars contribute at higher energies. In Figure~\ref{fig:PositronFraction}, the flux normalizations from individual 
pulsars are taken to be arbitrary. At lower energies many more pulsars contribute, thus features from individual pulsars 
get to be averaged out, leaving only the higher energy features present. Following and further refining the technique of 
Ref~\cite{Cholis:2017ccs}, on calculating a power spectrum from the positron fraction, i.e. performing an autocorrelation 
analysis, we search for such spectral features.

We use as a basis for the population of astrophysical sources the simulations of Ref.~\cite{Cholis:2021kqk}, that 
studied the properties of local Milky Way pulsars and how those can be constrained by high-energy cosmic-ray 
observations. Pulsars are known to be environments rich in electron-positron pairs. Those electrons and positrons 
can then be accelerated within the pulsar magnetosphere (see e.g. \cite{Guepin:2019fjb} for a recent update), as 
they cross the PWN or even the SNR shock front.  Depending on the exact region of particle 
acceleration, the spectra of electrons and positrons from pulsars may be equal or not. In this work, we will assume 
for simplicity that each pulsar injects into the ISM cosmic-ray electrons and positrons with the same spectrum. That 
assumes that the dominant high-energy acceleration takes place as these particles cross the PWN and the SNR fronts. 
As a result of this assumption of equal fluxes of high-energy positrons and electrons from individual pulsars, we 
can also predict that spectral features from individual sources will exist at same energy on both the positron 
and electron observed fluxes. Such coincident in energy features, can be searched for by cross-correlating the 
observed electron and positron fluxes. In this work, we show that the best way to perform such a cross-correlation
is to first evaluate and remove the overall smoothed electron and positron spectra, before performing such a 
cross-correlation analysis. We discuss the details of this technique and that of evaluating the power spectrum of 
the positron fraction in section~\ref{subsec:Data&Method}. 

For the groundwork of this study, we use the publicly available \textit{AMS-02} electron flux, positron flux and 
positron fraction data of Refs.~\cite{AMS:2019iwo, AMS:2019rhg} (see discussion in section~\ref{subsec:Data&Method}). 
We search for the presence of a power-spectral density that due to spectral features on the positron fraction is enhanced 
compared to regular noise expectations. We also search for a signal of coincident in energy features on the electron 
and positron fluxes. The methodology for the required 
techniques is discussed in section~\ref{subsec:Data&Method}, while our results are presented section
~\ref{sec:Results}. We test our findings against the library of Milky Way pulsar simulations 
of Ref.~\cite{Cholis:2021kqk}, to compare with the kind of power-spectrum and cross-correlation signals we could 
expect if pulsars are to explain the additional high-energy positrons. By performing a cross-correlation analysis using 
the \textit{AMS-02} electron and positron fluxes, we find clues of a positive correlation between these spectra, with
very similar characteristics to those expected from pulsars. We give our conclusions in section~\ref{sec:Conclusions}.

\vspace{-0.6cm}
\section{Data and Methodology}
\label{subsec:Data&Method}
\vspace{-0.2cm}
\subsection{Cosmic-ray observations}
\label{subsec:Data}
\vspace{-0.2cm}
We use observations by \textit{AMS-02} on the cosmic-ray positron fraction and positron and electron flux 
spectra spanning from 5 GeV to 1 TeV \cite{AMS:2019iwo, AMS:2019rhg}. While Refs.~\cite{AMS:2019iwo, 
AMS:2019rhg}, measure these spectra down to 0.5 GeV in energy, we ignore the 0.5 to 5 GeV range. 
These lower energies are dominated by uncertainties on the modeling of solar modulation that cosmic-rays 
undergo as they propagate inward through the heliosphere and also by the uncertainties relating to the 
production of electrons and positrons from inelastic collisions  of cosmic-ray nuclei with the ISM gas (see 
e.g. \cite{Orusa:2022pvp}). Moreover, at energies $<5$ GeV, we expect no spectral features relating 
to the presence of any individual Milky Way pulsar, as the 
number of contributing sources is too large and the resulting spectra from pulsar populations are very smooth \cite{Malyshev:2009tw, 
Profumo:2008ms, Grasso:2009ma, Cholis:2018izy, Orusa:2021tts, Cholis:2021kqk}. 

For our power-spectrum analysis, we rely on the positron fraction spectra from \textit{AMS-02} \cite{AMS:2019iwo}, 
that is shown in Fig.~\ref{fig:PositronFraction}. The positron fraction has smaller overall errors compared to the 
positron flux measurement of \cite{AMS:2019rhg}, as some of the systematic errors cancel out in taking a ratio 
of cosmic-ray fluxes. Instead, for our cross-correlation, we take the published  \textit{AMS-02}  electron and 
positron spectra of Ref.~\cite{AMS:2019iwo} and Ref.~\cite{AMS:2019rhg} respectively. These measurements 
are also shown in Fig.~\ref{fig:Electrons_and_positrons}. 
\vspace{-0.4cm}
\subsection{Methodology}
\label{sec:Method}
\vspace{-0.4cm}
The cosmic-ray electron and positron fluxes that we simulate have three components. The first, is the primary 
electrons accelerated by SNRs. The second one, is the secondary electrons and positrons produced at 
inelastic collisions in the ISM. Finally, local galactic pulsars produce cosmic-ray electrons and positrons.
\vspace{-0.6cm}
\subsubsection{Simulations of Milky Way pulsars}
\label{subsec:Pulsar_Simulations}
 \vspace{-0.2cm}
As a basis for the range of possibilities of spectral features from pulsars, we use the results of Ref.~\cite{Cholis:2021kqk} 
that produced a library of simulations on the Milky Way pulsar population. These simulations account for the discreteness 
of these sources both in position and age, uncertainties in the pulsar birth rate and the fact that each pulsar 
has a unique initial spin-down power, within a distribution that can be constrained by radio observations 
\cite{FaucherGiguere:2005ny, Manchester:2004bp, ATNFSite}. Moreover, these simulations account for uncertainties 
in the pulsars' time-evolution related to the pulsars' braking index and characteristic spin-down timescale. 
From each pulsar only a fraction $\eta$ of its rotational spin-down power gets converted to cosmic-ray electrons and 
positrons injected to the ISM. Moreover, each pulsar has a unique value on the conversion fraction $\eta$.  Uncertainties 
on that fraction $\eta$ and on the properties that describe the underlying 
distribution of $\eta$ for a population of pulsars have been included in the simulations of Ref.~\cite{Cholis:2021kqk} that we use.
Also, the injected electrons and positrons 
from each pulsar are described by a unique
injection spectral index $n$ within a range of allowed values. We assume that at injection the differential spectrum 
of cosmic rays is $dN/dE \propto E^{-n} \cdot exp\{ -E/ 10 \; \textrm{TeV}\}$. 
Finally, these simulations account for the fact that cosmic rays propagate through the ISM and the volume affected by the 
solar wind, the heliosphere. Uncertainties on the cosmic-ray propagation uncertainties through the ISM 
and the heliosphere are included. 

For the propagation through the ISM, the most relevant uncertainties are those of cosmic-ray diffusion and energy losses
 related to synchrotron radiation and inverse Compton scattering. Cosmic rays are assumed to propagate within   
a cylinder of radius 20 kpc and of half-height $z_{L}$ that is between 3 and 6 kpc. That cylinder has its center at the center
of the Milky Way. Diffusion is assumed to be homogeneous and isotropic, described by a diffusion coefficient  
$D(R) = D_{0} (R/(1 \; \textrm{GV}))^{\delta}$, where $\delta$ is between 0.33 and 0.5, with the limiting values describing 
 Kolmogorov and Kraichnan diffusion respectively \cite{1941DoSSR..30..301K, 1967PhFl...10.1417K, 1970PhFl...13...22K}. 
 At the cosmic-ray energies of interest, i.e. $\simeq$1GeV to 1 TeV, for the inverse Compton scattering the Thomson cross section \cite{1929ZPhy...52..853K} 
approximates well the Klein-Nishina one \cite{1970RvMP...42..237B}. As a result the electron/positron rate of energy losses 
can be simply written as $dE/dt = -b (E/(1 \, GeV ))^{2}$, where $b$ scales proportionally to the energy density in the cosmic 
microwave background photons, the interstellar radiation field photons and the local galactic magnetic field. \footnote{Such 
an approximation assumes the Thomson cross-section for the inverse Compton scattering and is valid for cosmic-ray electrons 
and positrons of energies up to a few 100s of GeV, but breaks down at energies closer to 1 TeV. Our analysis is less sensitive 
to these very high energies where still the statistical noise of the \textit{AMS-02} observations is very large.}

Our models 
also include the effects of diffusive reacceleration 
\cite{1994ApJ...431..705S} and convective winds in the ISM. All the relevant ISM propagation assumptions are tested to 
measurements by \textit{AMS-02} and \textit{Voyager 1} on the cosmic-ray hydrogen, and \textit{AMS-02} measurements of
the cosmic-ray spectra of helium, carbon and oxygen fluxes and the beryllium-to-carbon, boron-to-carbon and oxygen-to-carbon
ratio spectra studied in \cite{Cholis:2021rpp}. The properties of the ISM in those simulations are also in agreement
with results from Refs.~\cite{Trotta:2010mx, Pato:2010ih}. In addition, the electron and positron spectral measurements from \textit{AMS-02}, \textit{CALET} and
\textit{DAMPE} are used in~\cite{Cholis:2021kqk} to set constrains on the ISM local propagation. For the secondary fluxes \path{GALPROP} v54 
\cite{galprop, GALPROPSite}, has been used. Finally, to include the effects of cosmic-ray propagation through the heliosphere
which results in the solar modulation of all cosmic-ray spectra at energies bellow 50 GeV, we use 
the time-, charge- and rigidity-dependent formula for the solar modulation potential from \cite{Cholis:2015gna}, including recent 
analyses results from \cite{Cholis:2020tpi, Cholis:2022rwf}.

\begin{figure}
\begin{center}
\hspace{-0.5cm}
\includegraphics[width=3.7in,angle=0]{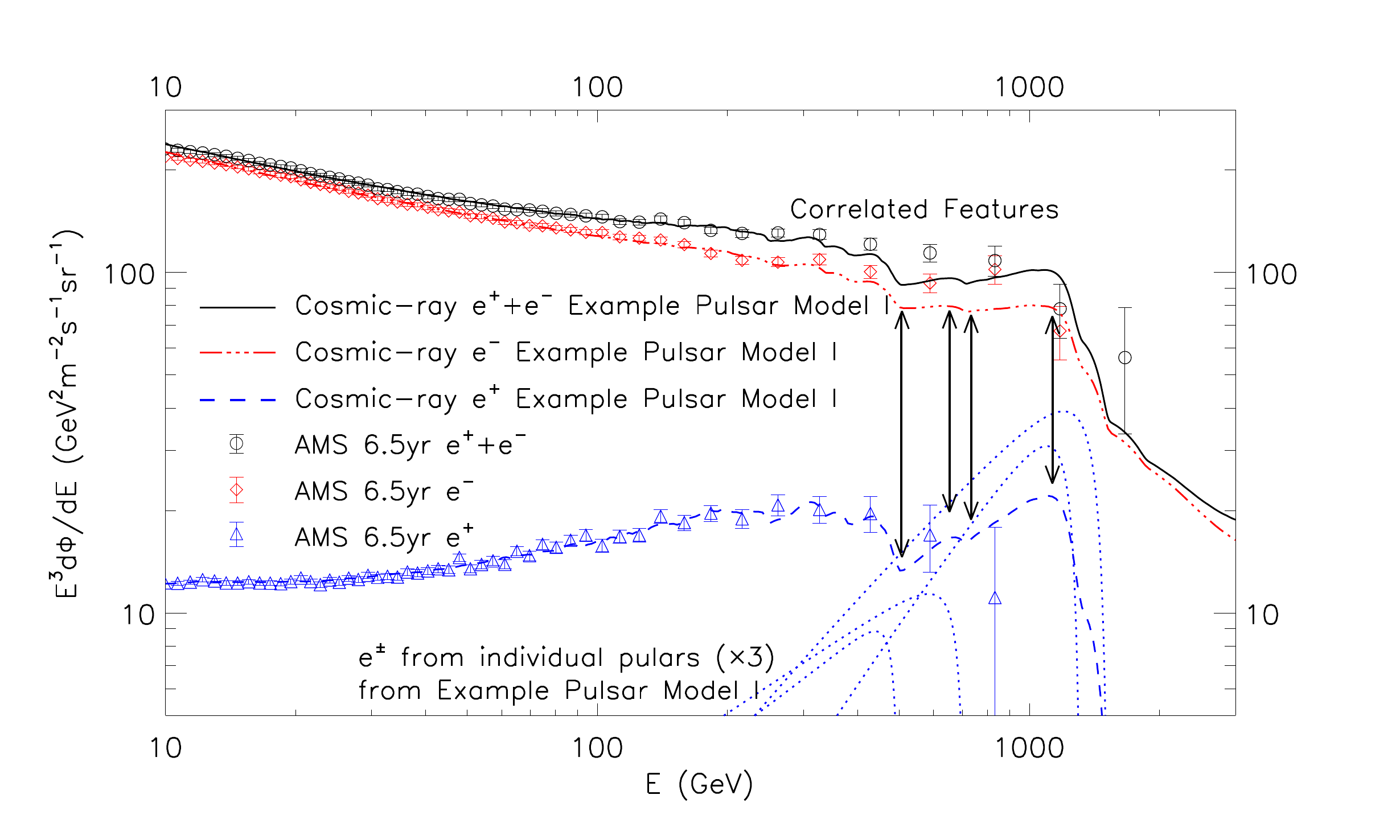}
\end{center}
\vspace{-0.6cm}
\caption{The predicted electron flux (red dashed-dotted line) and positron flux (blue dashed line) from an 
example pulsar population model. The \textit{AMS-02} electron and positron flux measurements are shown 
by the red and blue data points respectively. For a small number of the most prominent members of the pulsar population, we 
show in dotted blue lines the predicted electron and positron fluxes at Earth. In the black line and \textit{AMS-02} 
data points, we give the combined $e^{+} + e^{-}$ flux.}
\label{fig:Electrons_and_positrons}
\end{figure}

In Figure~\ref{fig:Electrons_and_positrons}, we give an example of the simulated electron cosmic-ray flux,
the positron cosmic-ray flux and the combined electron plus positron flux, from a population of pulsars including
primary and secondary cosmic-ray fluxes.
We also provide the relevant spectral data from \textit{AMS-02}. There are many more cosmic-ray electrons 
as the simulation includes the primary electron component from SNRs that exists only in electrons. Secondary fluxes 
are taken to be equal between electrons and positrons. For a few powerful and middle-aged to young pulsars (ages 
of $O(10^5)$ yr or younger), 
we also show their individual simulated fluxes. Pulsars are taken to contribute equal fluxes of cosmic-ray 
electrons and positrons. Powerful pulsars give spectral features to both the electron and the positron fluxes 
that coincide in energy. This is a property that we will use in our cross-correlation analysis.

As is showcased in Figs.~\ref{fig:PositronFraction} and~\ref{fig:Electrons_and_positrons}, spectral features 
from individual pulsars appear at energies above $\sim 50$ GeV \footnote{The stochastic nature of diffusion 
and ICS losses smoothens the cuspiness of spectral features \cite{Malyshev:2009tw, John:2022asa}. That most
dominantly affects the features at the lowest energies i.e. from the older pulsars. In our case, below 50 GeV we predict 
too many pulsars to observe any features from individual sources.}. In Ref.~\cite{Cholis:2021kqk}, about 
$7.3 \times 10^{3}$ unique Milky Way pulsar simulations were created to account for the combination of the
described above astrophysical uncertainties. Each of these simulations contained between $6\times 10^{3}$ 
and $20 \times 10^{3}$ unique pulsars within 4 kpc from the Sun, created over the last 10 Myr. Of these 
$7.3 \times 10^{3}$ local Milky Way pulsar population simulations, 567 simulations can fit the cosmic-ray 
positron fraction of \textit{AMS-02} above 15 GeV within 2$\sigma$ from a value of $\chi^{2}$/d.o.f. = 1.
There are 44 energy bins in the \textit{AMS-02} positron fraction measurement of Ref.~\cite{AMS:2019iwo}, 
between 15 and 1 TeV; and are fitted by seven free parameters. These parameters account for the 
uncertainties on the normalizations of primary cosmic rays, secondary cosmic rays and cosmic rays from 
pulsars. They account also for uncertainties on the injection spectral indices of the comic-ray primaries
and secondaries and on the modeling of solar modulation (see Ref.~\cite{Cholis:2021kqk} for more details). 
The simulations that are within 2$\sigma$ from an expectation of $\chi^{2}$/d.o.f. = 1, have a 
$\chi^{2}$/d.o.f.$ \leq 1.290$ on the positron fraction. In the remainder of this work, we use those 567 simulations 
as a basis on studying the kind of spectral features pulsar populations consistent with the current cosmic-ray 
measurements can give. We note that these 567 simulations are also within 2$\sigma$ from an expectation of 
$\chi^{2}$/d.o.f. = 1 to the positron flux measurement by \textit{AMS-02} \cite{AMS:2019rhg} and the electron 
plus positron measurements by \textit{AMS-02}, \textit{CALET} and \textit{DAMPE} 
\cite{AMS:2021nhj, Adriani:2018ktz, DAMPE:2017fbg}. The positron fraction combined statistical and systematic 
errors are the smallest among those observations and thus provide the main dataset in excluding Milky Way 
pulsar simulations. These simulations 
are used for both the power-spectrum analysis on the positron fraction and subsequently the cross-correlation 
of the predicted electron and positron fluxes. We also check the 
Milky Way pulsar simulations that were within 3$\sigma$ from an expectation of $\chi^{2}$/d.o.f. = 1, on the 
positron fraction, i.e. have a $\chi^{2}$/d.o.f.$ \leq 1.467$ and find that our results are qualitatively the same.
\vspace{-0.4cm}
\subsubsection{Power Spectrum on the Positron Fraction}
\label{subsec:PS_on_PF}
\vspace{-0.2cm}
While our simulations have underlying spectral features, statistical noise prominent in the high-energy 
cosmic-ray bins, will also cause fluctuations. That is true even if the underlying positron fraction spectrum 
is a smooth one, as the one depicted by the blue dashed line of Fig.~\ref{fig:PositronFraction}. We want 
to evaluate how often the signal from underlying high-energy spectral features is above the statistical noise. 
Building on the work of Refs.~\cite{Malyshev:2009tw, Cholis:2017ccs}, we want to use a power-spectrum 
analysis. We want to dissociate our work on the presence of spectral features in the cosmic-ray data, from any
particular energy bin. We do not know the properties of the entire local population of Milky Way pulsars 
-observing only a fraction of them through electromagnetic observations- nor the exact properties of the local 
ISM. Thus, we can not predict at what exact energies these spectral features will appear. Our analysis only relies
on the fact that a large fraction of the Milky Way pulsar simulations that we use predict some prominent 
features at the higher energies.

For each of the 567 pulsar astrophysical simulations, we generate 10 observational/mock realizations, i.e. 
we add noise on our simulations using the same energy bins and statistical $\%$ errors from Ref. 
\cite{AMS:2019iwo}. The noise-related fluctuations in some cases enhance and in other cases suppress 
the underlying spectral features. While the current measurement of the positron fraction seems (by eye) to 
suggest that the source of positrons with energies greater than $\simeq 500$ GeV is phasing out, statistically 
a positron fraction that has a plateau above 500 GeV, or even increases is still consistent with 
the data of~\cite{AMS:2019iwo}. We want to avoid any bias in our results from the large scale (in energy) evolution 
of the positron fraction. To do that, from each observational realization, we subtract its relevant smoothed spectrum and 
evaluate the power-spectral density (PSD) on its residual spectrum. For each positron fraction realization, its smoothed 
spectrum is evaluated by convolving with a gaussian function, whose width increases with energy. In doing 
that, we remove power from large scales in energy, i.e. low modes in the power spectrum. This is an important 
point, as in an actual analysis on the observed by \textit{AMS-02} positron fraction measurement, instrumental 
systematics that may span multiple energy bins are also removed in this manner. An example of such a systematic 
may be a misestimate of the instrument efficiency or cosmic-ray contamination. \footnote{Systematics 
affecting only few neighboring energy bins could still induce the small-scale fluctuations that 
we seek. In the future, it will be crucial that any correlated errors affecting only a few energy bins be well understood and 
accounted for.}
 
To evaluate the PSD on the residual positron fraction spectrum, we take as the equivalent of the ``time'' 
parameter to be $ln(E/\textrm{GeV})$. Approximately the energy binning of the \textit{AMS-02} data is done in equal 
intervals of $ln(E)$ up to $\sim 200$ GeV. For this analysis where we use observational/mock realizations, 
we take the energy binning to be on exactly equal bins in the $ln(E)$-space. The specific logarithmic energy 
binning we assume is,
\begin{equation} 
ln(E_{i}/\textrm{GeV}) \equiv x_{i} = x_{0} + a \cdot i, 
\label{eq:EbinSim}
\end{equation}
with $x_{0} =1.6564$ (5.24 GeV) and $a=0.077$. This is a slightly more dense energy binning than \textit{AMS-02},
but only at the higher energies. Such a binning may be used in measurements that include longer observation 
times than the currently published 6.5-year ones.  When comparing to current data, we go up to $i=59$ (492 GeV). 
We calculate the PSDs for each of the 567$\times$10 observational/mock realizations that include noise, giving  
a scatter on the PSDs of these realizations evaluated for any given underlying astrophysical simulation.

We need to compare the PSDs of the 5670 observational realizations from our pulsar population simulations to 
what the PSDs are expected to be by just having noise on the residual positron fraction spectra. To do that, we 
take the \textit{AMS-02} measurement of the positron fraction, and convolve it with the same gaussian function 
used to derive the smoothed positron fraction simulated spectra for the 567 pulsar population simulations. That 
gives us the smooth \textit{AMS-02} positron fraction spectrum. We then use that smooth spectrum and add noise on it, 
following the same procedure as for the pulsar population simulations. We create 1000 simulated \textit{AMS-02} 
positron fraction measurements. From each of those simulated measurements, we subtract the smooth 
\textit{AMS-02} positron fraction spectrum and then on these residuals we evaluate 1000 PSDs using the same 
binning of Eq.~\ref{eq:EbinSim}. These 1000 PSDs give us the PSD-ranges due to the statistical noise. 

\begin{figure}
\begin{center}
\vspace{-0.4cm}
\hspace{-0.5cm}
\includegraphics[width=3.7in,angle=0]{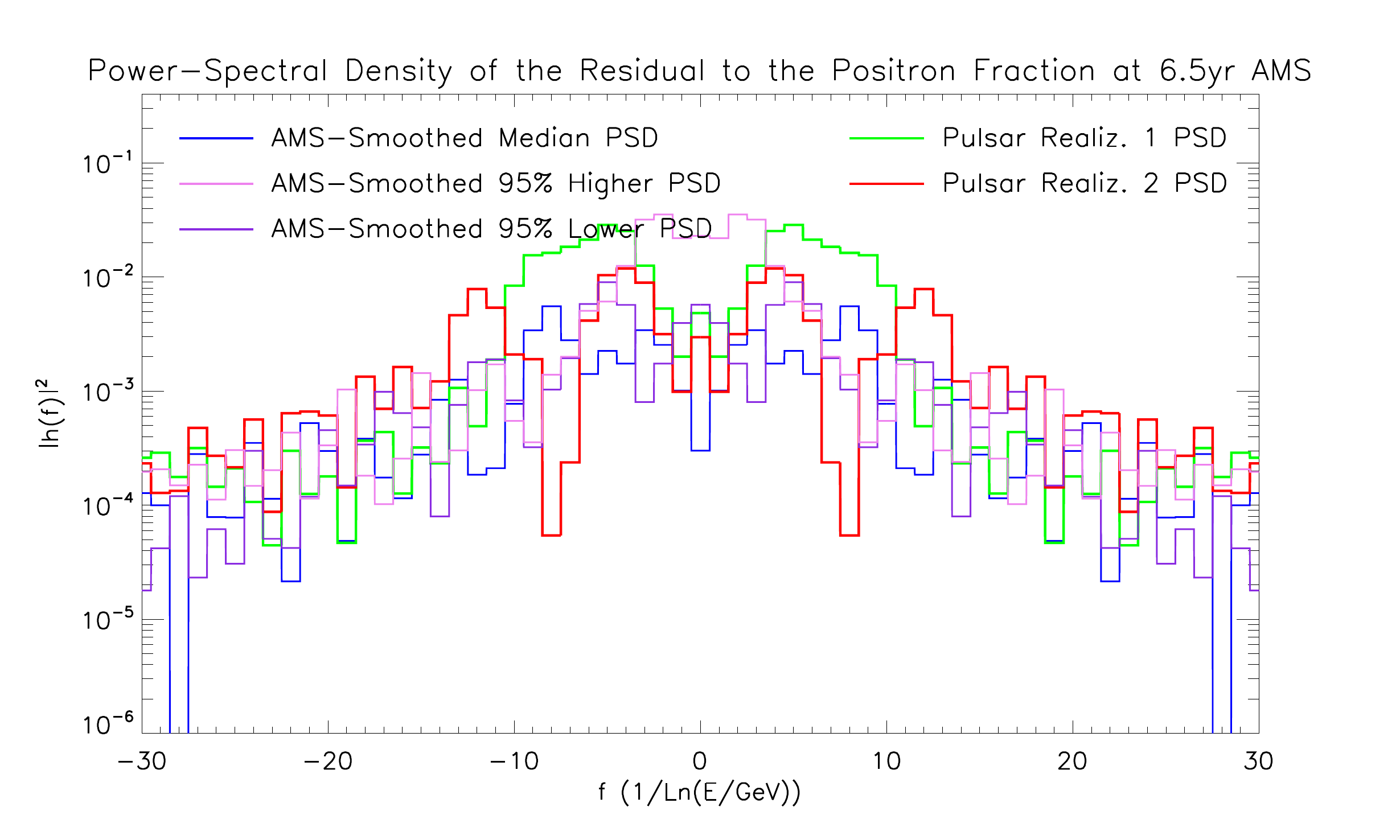}
\end{center}
\vspace{-0.6cm}
\caption{The power-spectral density of the residual cosmic ray positron fraction as a function of frequency, 
where frequency is taken to be $1/ln(E/(1 \; \textrm{GeV}))$. Assuming that the positron fraction is a smooth 
featureless function with fluctuations arising due to noise, we give the expected PSD as a fraction of frequency. 
That is shown in 
the blue, pink and purple lines for the simulations with the median total power in the PSD, for the 95-percentile (95$\%$) higher total power in the PSD and for the $95\%$ lower total power in the PSD respectively (see text for details). In the green and red 
lines we showcase that if the pulsars have inherent spectral features, those will remain as fluctuations 
in the residual positron fraction spectrum and even with noise present, can give an enhanced PSD in a certain 
frequency range (modes 3-13 in these examples).}
\label{fig:PSD_vs_frequency}
\end{figure}

In Fig.~\ref{fig:PSD_vs_frequency}, we plot three PSDs of the residual to the simulated positron fraction, using the 
6.5-year data for the noise~\cite{AMS:2019iwo}. Assuming that the positron fraction is a smooth featureless 
function (as the one shown in Fig.~\ref{fig:PositronFraction} by the blue dashed line), with only noise causing 
fluctuations around it, we plot the expected PSD $|h(f)|^{2}$ as a faction of ``frequency'' $f = 1/ln(E/\textrm{GeV})$. 
Ranking the 1000 \textit{AMS-02} realizations based on the total power i.e. $\Sigma_{i=-30}^{i=+30} |h(f_{i})|^{2}$, we 
can identify the \textit{AMS-02} realization that has a median value on its total power. This is given by the blue line in 
Fig.~\ref{fig:PSD_vs_frequency}. By the same ranking, we identify the \textit{AMS-02} realization that is at the 
95-percentile higher range in terms of its total power (95$\%$ higher PSD), given by the pink line and the realization 
that is at the 95-percentile lower range in terms of its total power (95$\%$ lower PSD), given by the purple line. Finally,
we plot the PSDs evaluated from for two pulsar populations (red and green lines). We noticed that a significant fraction
of our pulsar populations simulations have an enhanced PSD in modes 3-13, compared to the noise PSD ranges, 
which we will explore in more detail in section~\ref{sec:Results}. 

In addition, for every one of the 60 modes that we use, we rank the 1000 coefficients from the 1000 \textit{AMS-02} 
noise realizations. We use the 68$\%$ ranges to derive the $1 \sigma$ error-bars per mode.  We note that we do not 
expect any correlations between modes. We can use those $1 \sigma$ error-bars per PSD mode, to calculate a 
$\chi^{2}$ fit on each of the PSDs derived from our pulsar populations. Like with the total power we can rank the 
1000 mock realizations of the \textit{AMS-02} smooth parametrization in terms of the $\chi^{2}$ quality fit and 
present the median among those 1000 realizations. We can do the same to get the $68\%$, $90\%$ and
$98\%$ ranges for the 1000 mock realizations of the \textit{AMS-02} with noise. We use these ranges again to 
compare in terms of their $\chi^{2}$ quality, the expected PSDs of the pulsar population astrophysical realizations 
to the PSDs we get just due to statistical noise. This is an alternative way to check if pulsar related features can give 
a signal in the PSD that is above the noise.

\subsubsection{Cross-correlating the Electron and Positron Spectra}
\label{subsec:CC_on_EP}

Like with the power-spectrum analysis, we are focused in comparing the small-scale features of the electron and 
positron fluxes coming from pulsars to random noise fluctuations. To study these features, rather than the full electron 
and positron flux spectra, we need to evaluate 
the residual spectra after subtracting a smooth function that describes well the large scale energy-dependence of these
spectra. The smoothed models that we use are crucial to the analysis. 

\begin{figure}
\begin{center}
\hspace{-0.5cm}
\includegraphics[width=3.7in,angle=0]{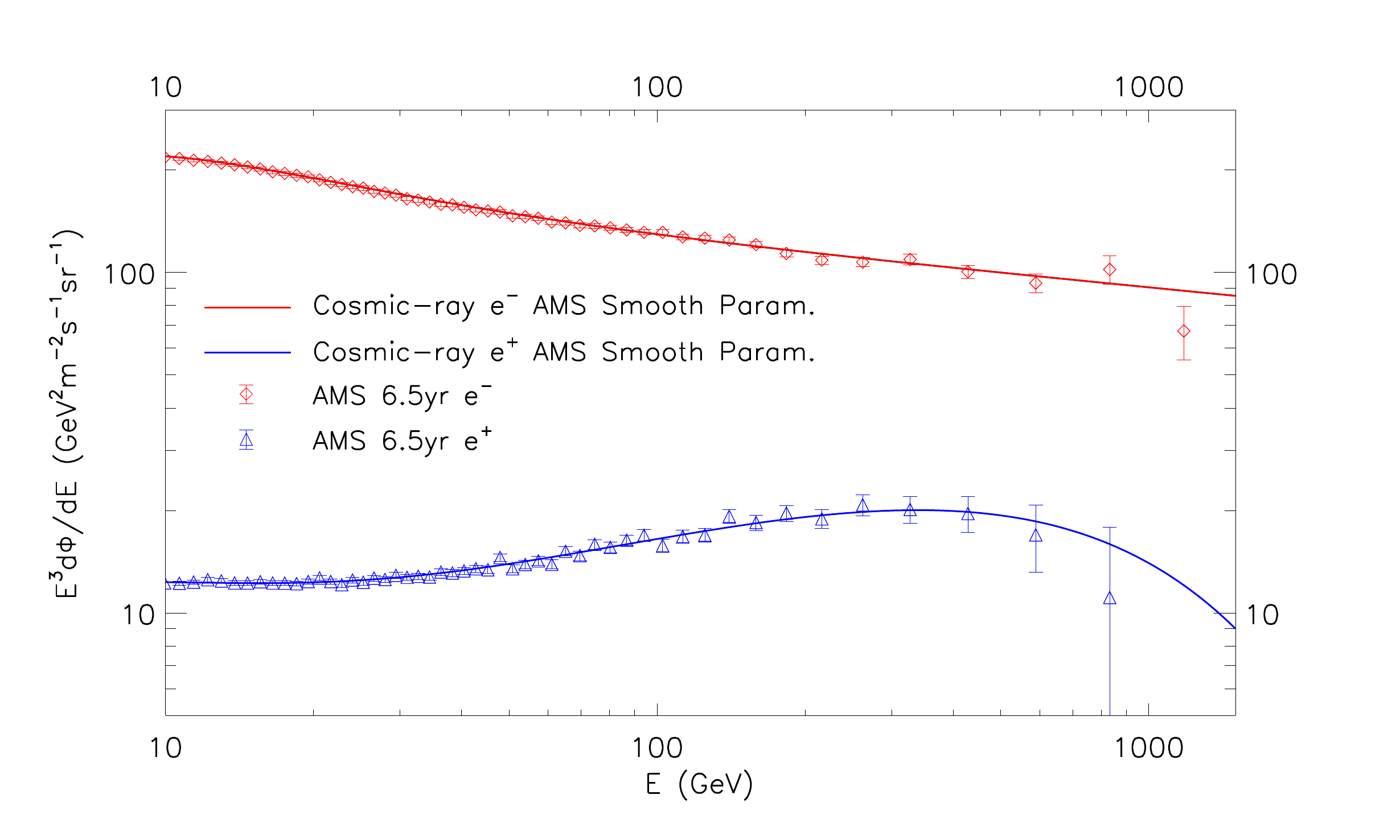} \\
\vspace{-0.6cm}
\includegraphics[width=3.7in,angle=0]{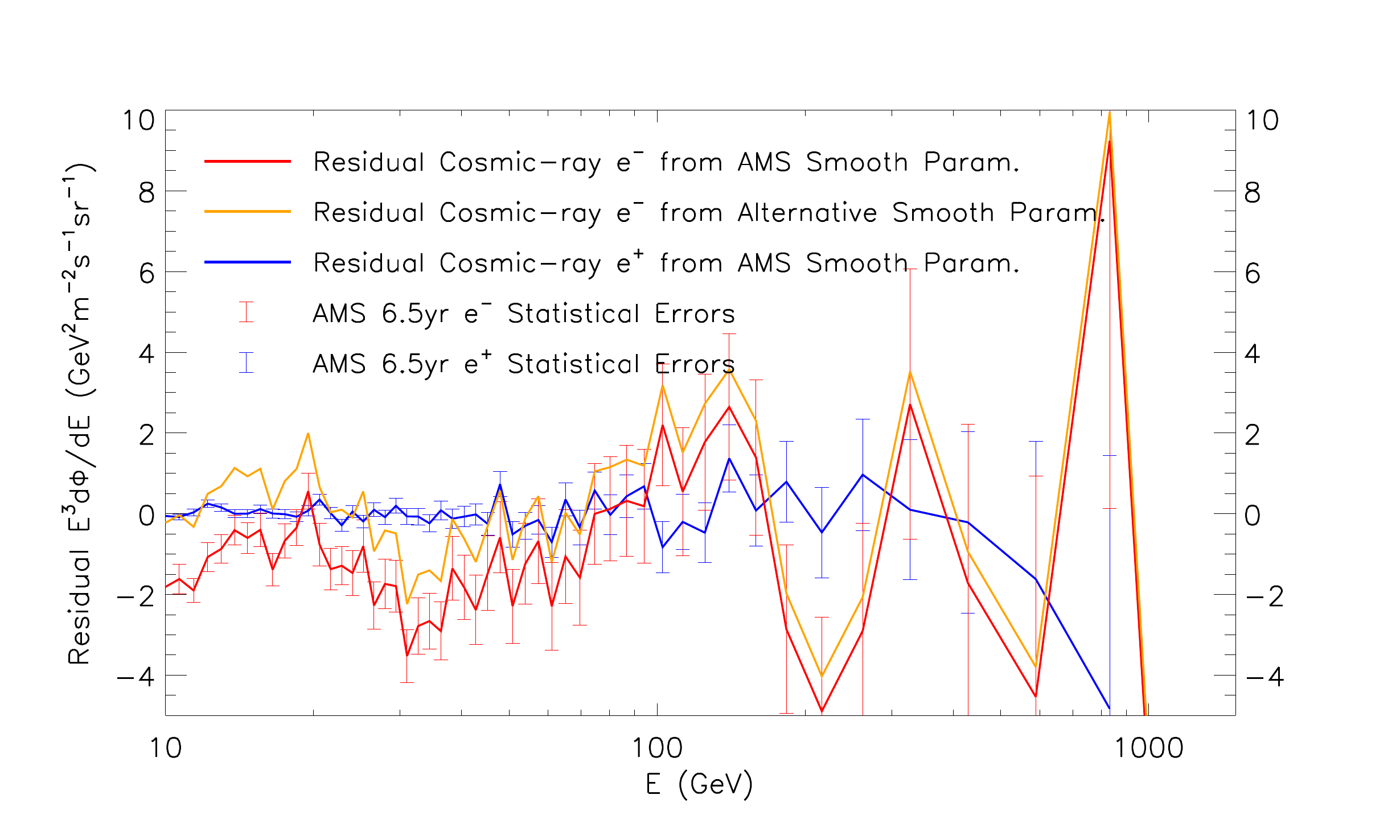}
\end{center}
\vspace{-0.6cm}
\caption{Top panel: the \textit{AMS-02} electron (red) and positron (blue) flux measurements and their respective 
smooth parameterizations from Refs.~\cite{AMS:2019iwo, AMS:2019rhg}. Bottom panel: the residual electron and 
positron fluxes. For the electrons we derive an alternative smooth parameterization (orange) that allows us to then 
study the residual spectra upon which our cross-correlation analysis is performed. We also show the statistical errors
for the electron and positron measurements after 6.5 years. We do not over-plot statistical errors for the alternative 
smooth spectrum electron parameterization as they are the same in size as the red ones.}
\label{fig:ElecPos_Residuals}
\end{figure}

The \textit{AMS-02} collaboration has provided 
a model/parameterization for each of the electron and positron spectra \cite{AMS:2019iwo, AMS:2019rhg}. While their 
parameterization works for the positron spectrum, it overpredicts the electron observations at the lower energies as 
shown in the residuals in Fig.~\ref{fig:ElecPos_Residuals}. We focus on energies above 10 GeV as below 
that energy, features from individual pulsars are not expected. In Fig.~\ref{fig:ElecPos_Residuals}, in the top panel 
we show the positron and electron measurements above 10 GeV as well as the parameterizations provided by the 
\textit{AMS-02} collaboration. As we show in the bottom panel of Fig.~\ref{fig:ElecPos_Residuals}, (blue line) once 
removing from the \textit{AMS-02} data the smooth parameterization, the resulting residuals span only a small number of 
energy bins. The most prominent remaining features in positrons are seen around 12 GeV and around 21 GeV. Those 
may be related to physical processes from further away distances than few kpc 
away from the Sun as suggested recently in \cite{Cholis:2022ajr}. However, in the electrons we don't see as prominent 
features \footnote{As the electron flux at the 10-20 GeV energy is more than 10 times larger to the positron flux, an additional 
electron/positron component giving a feature at the positron spectrum might be very difficult to identify at the electron spectrum.}. 
In the bottom panel of Fig.~\ref{fig:ElecPos_Residuals},  we show (in red) the residual cosmic-ray electron 
flux once removing the smooth spectrum as parameterized by \textit{AMS-02}. The residual flux systematically underpredicts 
the observed flux between 20 
and 80 GeV even if its overall quality of fit has a $\chi^{2}/\textrm{d.o.f.} <1$. For that reason we performed a $\chi^{2}$-fit,
deriving slightly different values of the parameters within the same parametrization given in Table~\ref{tab:ElectronParam}. 
Our choice fits the overall 
spectra better and gives residuals that are more localized in energy. The difference between the \textit{AMS-02} 
parameterization and the alternative one is too small to show up in the top panel of Fig.~\ref{fig:ElecPos_Residuals},
but can be seen in the bottom panel. The orange line does fluctuate around zero, however in electrons with
energy less than 40 GeV there are still neighboring bins that may be correlated. 

Here, for clarity on the assumptions that we use, we repeat what the parameterizations from Refs.~\cite{AMS:2019iwo, 
AMS:2019rhg} are. The positrons smooth spectrum is given by,
\begin{eqnarray}
\frac{d\Phi_{e^{+}}}{dE}(E) &=& \frac{E^{2}}{\hat{E}^{2}}  [ C_{d} (\hat{E}/E_{1} )^{\gamma_{d}}  \nonumber \\
                         &&+ \; C_{s} (\hat{E}/E_{2})^{\gamma_{s}}  \cdot exp\{ - \hat{E}/E_{s}\} ].
  \label{eq:Positrons_Smooth_func}
\end{eqnarray}
The parameter $\hat{E}$ is energy dependent and equal to  $\hat{E} = E + \phi_{e^{+}}$, where $\phi_{e^{+}} = 1.10$ GeV.
The other parameters of Eq.~\ref{eq:Positrons_Smooth_func}, are $C_{d} = 6.51 \times 10^{-2} [\textrm{m}^{2}\, \textrm{s} \,\textrm{sr} \, \textrm{GeV}]^{-1}$, 
$\gamma_{d}=-4.07$, $C_{s} = 6.80 \times 10^{-5} [\textrm{m}^{2}\, \textrm{s} \,\textrm{sr} \, \textrm{GeV}]^{-1}$, 
$\gamma_{2}=-2.58$, $E_{1} = 7.0$ GeV, $E_{2} = 60.0$ GeV and $E_{s} = 810$ GeV.
 

The electrons smooth spectrum is given by,
\begin{eqnarray}
\frac{d\Phi_{e^{-}}(E)}{dE} &=& \frac{E^{2}}{\hat{E}^{2}}  \left[ 1+ \left(\frac{\hat{E}}{E_{t}} \right)^{\Delta \gamma_{t}} \right]^{-1} \nonumber \\
                         &\times& \left[ C_{a} \left(\frac{\hat{E}}{E_{a}} \right)^{\gamma_{a}}  + C_{b} \left(\frac{\hat{E}}{E_{b}} \right)^{\gamma_{b}} \right].
  \label{eq:Electrons_Smooth_func}
\end{eqnarray}
The parameter $\hat{E}$ is also energy dependent and equal to  $\hat{E} = E + \phi_{e^{-}}$. In Table~\ref{tab:ElectronParam}, we 
give the values of the parameters in Eq.~\ref{eq:Electrons_Smooth_func} for the \textit{AMS-02} and our alternative parameterizations.
The values for $E_{a}$ and $E_{b}$ are fixed in both the results of~\cite{AMS:2019iwo} and our analysis. We created a $\chi^{2}$ 
per degree of freedom function and optimized it with dual annealing. This method was chosen as it can fit several parameters at 
once and the function need not be linear to use it. This method is stochastic, thus, each time it runs, we receive slightly different
 values for these parameters. 
\begin{table}[t]
    \begin{tabular}{ccc}
         \hline
            Parameter & \textit{AMS-02} value & Alternative value  \\
            \hline \hline
            $\phi_{e^{-}}$ (GeV) & 0.87 & 0.87 \\
            $E_{t}$ (GeV) & 3.94 & 3.94 \\
            $\Delta\gamma_{t}$ & -2.14 & -2.15 \\
            $C_{a}$ ($[\textrm{m}^{2}\, \textrm{s} \,\textrm{sr} \, \textrm{GeV}]^{-1}$)  & $1.13\times 10^{-2}$ &  $1.12\times 10^{-2}$ \\     
            $E_{a}$ (GeV) & 20 & 20 \\
            $\gamma_{a}$  & -4.31 & -4.31 \\
            $C_{b}$ ($[\textrm{m}^{2}\, \textrm{s} \,\textrm{sr} \, \textrm{GeV}]^{-1}$)  & $3.96\times 10^{-6}$ &  $3.93\times 10^{-6}$ \\
            $E_{b}$ (GeV) & 300 & 300 \\
            $\gamma_{b}$ & -3.14 & -3.14 \\  
            \hline \hline 
        \end{tabular}
    \caption{The parameters describing the smooth function for the cosmic-ray electron flux in Eq.~\ref{eq:Electrons_Smooth_func}. 
    In the second column we repeat the information from Ref.~\cite{AMS:2019iwo}.}
    \label{tab:ElectronParam}
\end{table} 

Under the original parameters, for energies above 5 GeV we get a $\chi^{2}$/d.o.f. of 0.65 for the electron data once adding in quadrature the statistical 
and systematic errors. Under the new parameters, for the same energy range, we get instead a $\chi^{2}$/d.o.f. of 0.37. We note 
however, that on the cross-correlation analysis we only retain the statistical errors that are shown in Fig.~\ref{fig:ElecPos_Residuals}
(bottom panel). The systematic/instrumental errors span several energy bins, unlike the spectral features we are searching for and 
thus can be ignored for the cross-correlation purposes.

The cross-correlation function \cite{Kido:2015}, is an operation which takes discrete functions $x(n)$ and $y(n-m)$, 
and creates a function $r_{xy}(m)$ which describes the relatedness of each point $n$ as a function of a shift variable $m$. 
In our specific case we define a similar type of cross-correlation function $r_{xy}(m)$ described by, 
\begin{eqnarray}
r_{xy}(m) =  \frac{1}{L-|m|} \begin{cases} \Sigma_{n=m}^{L-1} \frac{x(n) y(n-m)}{\sigma_{x}(n) \sigma_{y}(n-m)} \; \; \textrm{for $m \ge 0$} \\
\Sigma_{n=0}^{L-1+m} \frac{x(n) y(n-m)}{\sigma_{x}(n) \sigma_{y}(n-m)}   \; \; \textrm{for $m<0$}
\end{cases}
  \label{eq:CC_formula}
\end{eqnarray}
$L$ is the number of discrete data points for which we know $x$ and $y$ functions; while $\sigma_{x}$ and $\sigma_{y}$ are
the respective errors (standard deviations). In our case 
$L=50$ starting our
analysis above 10 GeV and going up to the last data point in positrons (700-1000 GeV). $x(n)$ and $y(n)$ are respectively 
the positron and electron residual spectra shown in  Fig.~\ref{fig:ElecPos_Residuals} (bottom panel) i.e. $E^{3} d\Phi(E)/dE$ (in $
[\textrm{m}^{2}\, \textrm{s} \,\textrm{sr}]^{-1}  \textrm{GeV}^2$). $\sigma_{x}$ and $\sigma_{y}$ the respective systematic errors. 
The parameter $n$ is set to 0 for the bin centered at 10.67 GeV, covering the cosmic rays with energy of 10.32 to 11.04 GeV. The maximum range of energies tested is from 10 GeV and up to 1.0 TeV. A positive shift $m$, represents shifting the electron 
flux to lower energies.  We test up to 
values of $|m|=24$, however the large shifts suffer from noise as the number of bins drops down to half the original number.
We do not show results for larger shifts as there is no physical reason to see any correlation between bins separated many 10s 
to 100s of GeV from each other.

We note, \textit{an important caveat} about the specific version of Eq.~\ref{eq:CC_formula} that we use. Conventional 
cross-correlation would require in the fraction of Eq.~\ref{eq:CC_formula}, to have instead of 
$x(n) y(n-m)/(\sigma_{x}(n) \sigma_{y}(n-m))$, the fraction
$(x(n)-\bar{x})(y(n-m)-\bar{y})/(\sigma_{x}(n) \sigma_{y}(n-m))$. That would give a well-defined cross-correlation that  
would give correlation coefficients $r_{xy}(m)$ within $[-1,+1]$. However, that would require to evaluate averages for $x(n)$
and $y(n)$, where different values of $n$ are not different instances of a time-series but different energies, where different 
physical phenomena occur. Pulsars are most dominant at high energies while secondary cosmic rays are more important at the lower energies. Our Eq.~\ref{eq:CC_formula} uses the residual spectra that still may have $\bar{x} \neq 0$ and $\bar{y} \neq 0$. 
Our residual spectra may still contain features spanning a few bins as a pulsar population may predict pulsars giving partially 
overlapping spectral features. If that is the case, we don't want this information to be completely lost. By subtracting from a 
smooth function that describes the overall spectral evolution, \textit{while} not removing the average of the entire 10-1000 GeV 
range we achieve that goal. Also, removing the $\bar{x}$ and $\bar{y}$ would associate in the residual spectra physical conditions at 
10 GeV to 1000 GeV and as at higher energies the noise is very large would make our results overly sensitive to the exact 
measurements at the highest energy bins. Thus, formally our Eq.~\ref{eq:CC_formula}, can give a correlation 
coefficient $r_{xy}(m)$ that takes values beyond the range of $[-1,+1]$. We also note that the $\sigma_{x}$ and $\sigma_{y}$ 
are not the standard deviations evaluated from all the $x$ and $y$ measurements. Each point $x(n)$ and $y(n)$ has its 
own uncertainty, directly related to the statistical noise of \textit{AMS-02} at that energy bin, i.e. to the number of positron 
and electron cosmic rays detected within
a specific energy range. As the \textit{AMS-02} keeps making observations, these statistical errors will become smaller. If 
there is an underlying correlation signal this will show up by an increasing $r_{xy}(m)$ with observation time around a specific range of
$m$ values. 

In practice, for 
our pulsar population simulations the chance of getting a $r_{xy}(m)$ beyond the range of $[-1,+1]$ is extremely small and when 
it happens it has to do with difficulty in evaluating the proper residual spectra. However, in the case of the actual 
\textit{AMS-02} measurements, evaluating a proper residual spectrum is still a challenge at the lower electron energies.

\begin{figure}
\begin{center}
\hspace{-0.5cm}
\includegraphics[width=3.7in,angle=0]{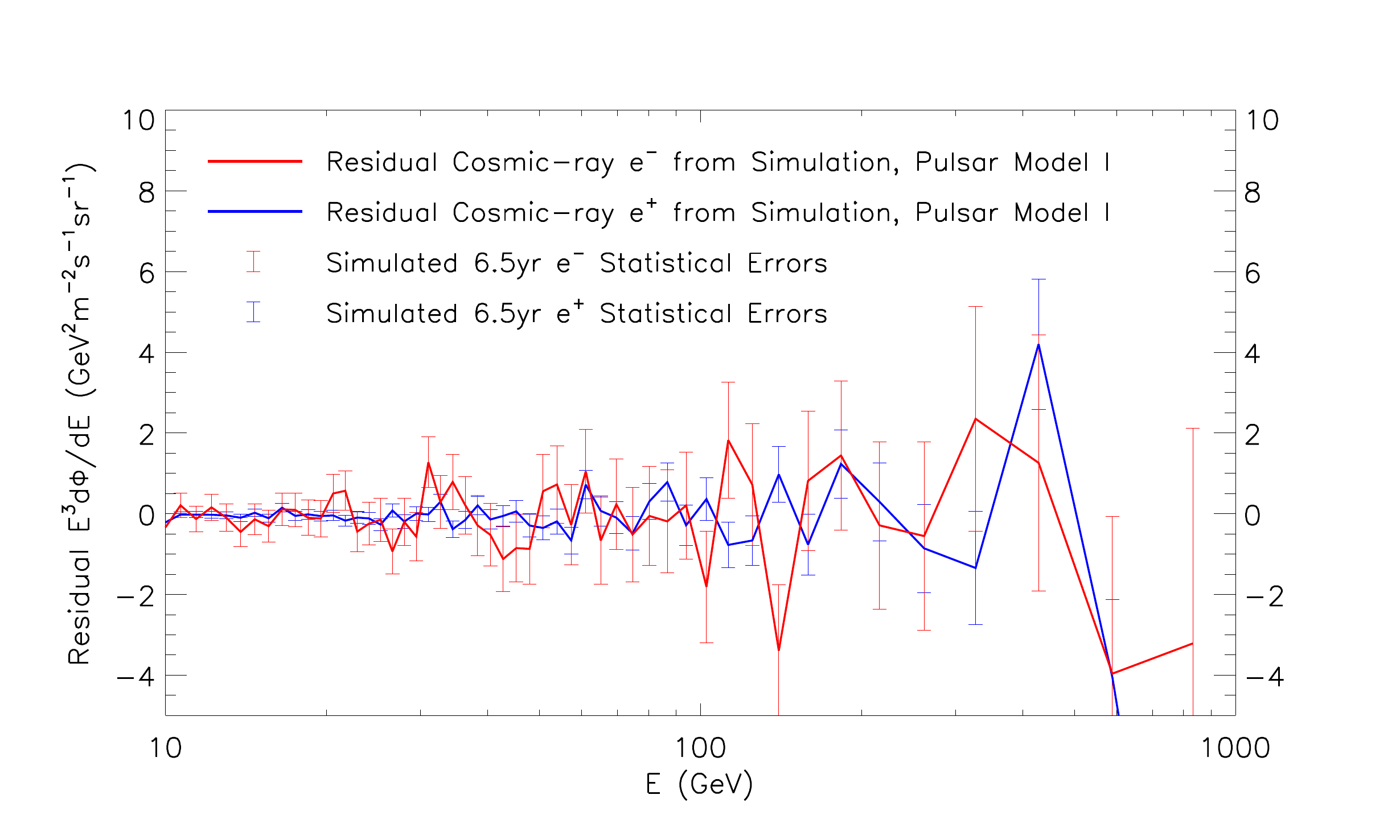} \\
\vspace{-0.2cm}
\includegraphics[width=3.7in,angle=0]{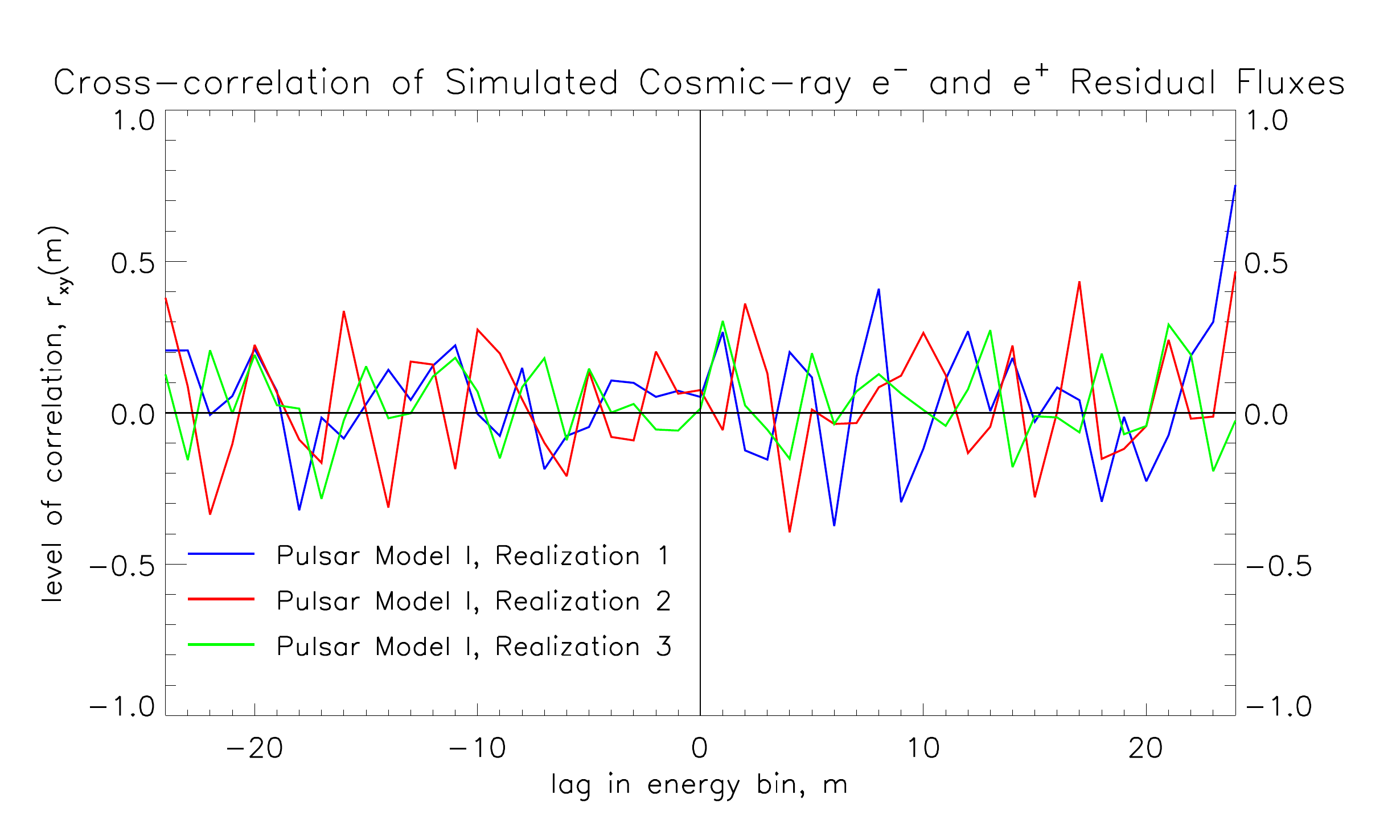}
\end{center}
\vspace{-0.6cm}
\caption{Top panel: like with Fig.~\ref{fig:ElecPos_Residuals} (bottom panel), the residual electron (red) and positron (blue) 
fluxes from one realization of the Milky Way pulsar population Model I. The statistical errors are simulated for 6.5 years of 
\textit{AMS-02} measurements. Bottom panel: the cross-correlation function between the electron and positron residual 
fluxes for three different observational realizations of the same underlying pulsar population (Model I). Pulsars give a peak of
cross-correlation signal around $m=0$ to +1.}
\label{fig:ElecPos_Residuals_PulsarSim}
\end{figure}

The dependence of the $r_{xy}(m)$ correlation coefficient as a function of the shift $m$ is what we study here. To see 
if there can be a cross-correlation signal related to pulsars in the \textit{AMS-02} measurements, we need to compare to the 
expectations from our pulsar population simulations. In Fig.~\ref{fig:ElecPos_Residuals_PulsarSim}, we show, as an example the residual
electron and positron spectra from one observational realization of a Milky Way pulsar population,``Model I'' simulation. We remind 
the reader that we create 10 observational realizations per Milky Way pulsar population simulation. These observational
realizations contain the information of statistical noise for 6.5 years of \textit{AMS-02} observations. For each of them, we then 
evaluate the smoothed electron and positron spectra by convolving the simulated spectra with the same kind of gaussian 
function described in~\ref{subsec:PS_on_PF}. Then, we subtract from the simulated realization electron and positron spectra 
their equivalent smoothed spectra, getting
the residual electron and positron spectra. It is these residual spectra that we cross-correlate. In the bottom panel of Fig.~\ref{fig:ElecPos_Residuals_PulsarSim}, we show the cross-correlation coefficient $r_{xy}(m)$ as a function of $m$, for 
three observational realizations of the same underlying pulsar population ``Model I'' simulation. The blue line on the bottom panel, 
is the one calculated from cross-correlating the residual spectra of the top panel on Fig.~\ref{fig:ElecPos_Residuals_PulsarSim}.
For the cross-correlation, we use the same energy binning as the \textit{AMS-02} results of \cite{AMS:2019iwo, AMS:2019rhg}. 
We do not need the energy bins to be equally separated (in $log(E)$) as we need for the power-spectrum analysis of
section~\ref{subsec:PS_on_PF}.
We note that a peak of cross-correlation coefficient occurs around $m$=0 or +1, typically being between $0$ and $+3$. A positive
$m$ value for the peak suggests that the pulsar features on the electrons typically appear to be shifted by one bin at lower energies. 
In section~\ref{sec:Results}, we describe these and other properties of the cross-correlation signals we expect from pulsar
populations for the entire set of simulations.

\section{Results}
\label{sec:Results}

\subsection{Power-Spectrum Analysis on Pulsar Population Simulations}
\label{subsec:Result_PS_on_PF}

\begin{figure}
\begin{center}
\vspace{0.4cm}
\hspace{-0.15in}
\includegraphics[width=3.52in,angle=0]{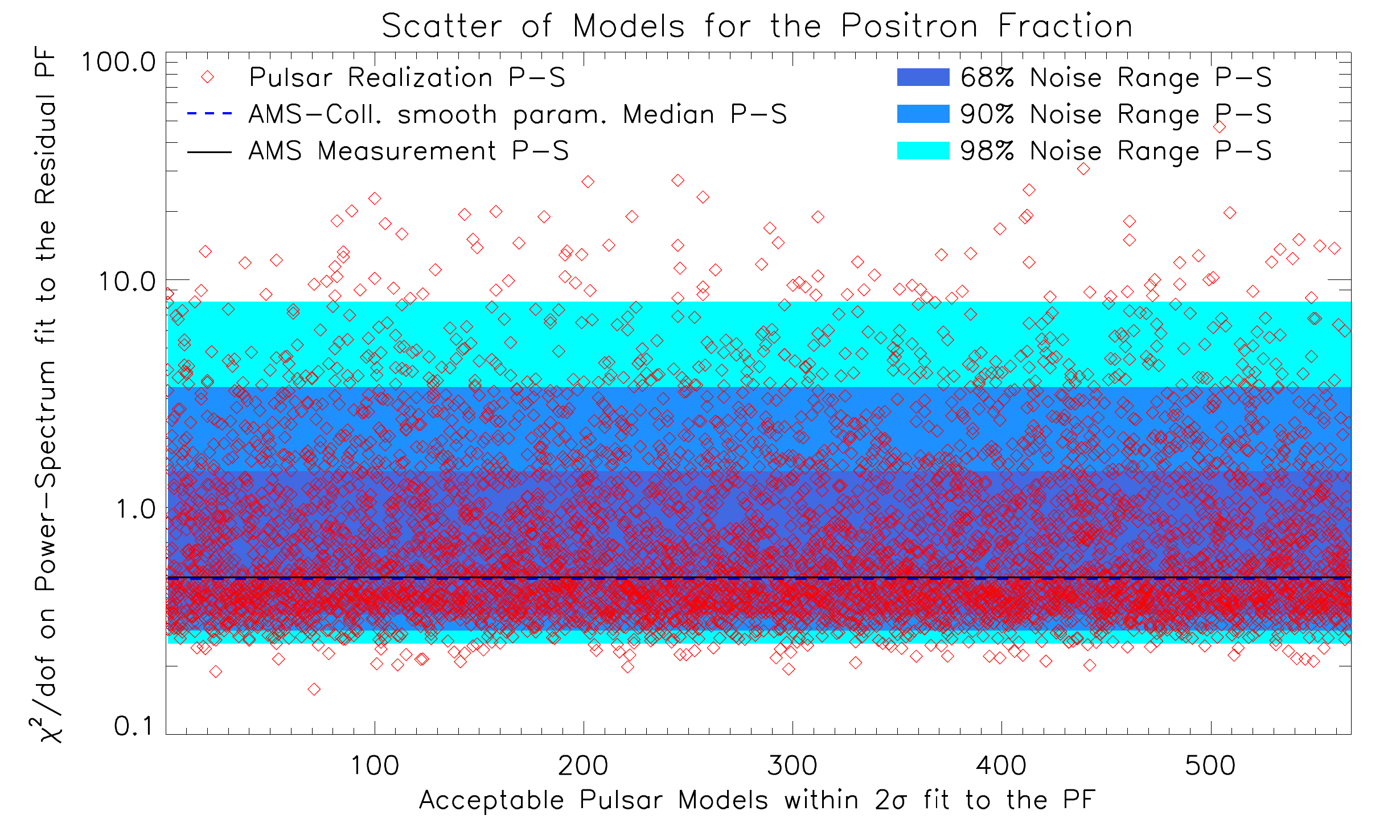}
\end{center}
\vspace{-0.5cm}
\caption{The scatter of the simulated Milky Way pulsars observational realizations in the PSD $\chi^{2}$/d.o.f., evaluated 
from their residual
positron fraction spectrum. We plot 10 realizations (in red diamonds) for each of the 567 pulsar population astrophysical simulations
that are consistent within $2\sigma$ to the positron fraction measurement (see section~\ref{subsec:Pulsar_Simulations} 
for details). The blue bands include the noise ranges for the PSD $\chi^{2}$/d.o.f.. The black line gives the PSD of the 
\textit{AMS-02} measurement after 6.5 years. The black line (measurement) overlaps very well with the median expectation 
of $\chi^{2}$/d.o.f. coming from a smooth spectrum with noise added (blue dashed line). This shows no evidence for 
features within the 6.5-year data.}
\label{fig:AutoCorrResults_chi2}
\end{figure}

We start with the results of the power-spectrum analysis on the pulsar population simulations. As we described in 
section~\ref{subsec:PS_on_PF}, using the 1000 \textit{AMS-02} mock positron fraction simulations we can 
calculate $1 \sigma$ error-bars per each of the 60 PSD modes.  From them, we can calculate the 
$\chi^{2}$ fit on each of the pulsars PSDs. In Fig.~\ref{fig:AutoCorrResults_chi2}, we show the 
PSD $\chi^{2}$-distribution (red diamonds) from each of the 10 observational realizations of our 567 
pulsar population simulations. Each pulsar population simulation is in a different position on the $x$-axis.
These are ranked from left to right starting from the model that provides the best fit the positron fraction 
spectrum, with all of them being within a 2$\sigma$ from an expectation of $\chi^{2}$/d.o.f. = 1 to the positron 
fraction (and flux) measurement. Given that we rank our models from better to worse fit, there is no clear pattern 
between the quality of fit that models provide to the observed positron fraction spectrum and their 
respective PSD $\chi^{2}$/d.o.f.. The 1000 mock realizations of the \textit{AMS-02} smooth 
parametrization for the positron fraction, can be ranked in terms of their  PSD $\chi^{2}$/d.o.f. (our $y$-axis). 
From there we get the respective $68\%$, $90\%$ and $98\%$ (two-sided) ranges, for the case where only noise is 
present, i.e. no underlying small-scale features. These are shown in the three different shades of blue in 
Fig.~\ref{fig:AutoCorrResults_chi2}. By comparing the red diamonds to the blue ranges, we notice that 
pulsar population simulations have a tendency for larger values of PSD $\chi^{2}$s.  We find that with the 6.5-years 
of sensitivity 1.8$\%$ (7.2$\%$) of the 5670 observation realizations lie outside the $99 \%$ ($95 \%$) upper 
band end, i.e. above $y$-axis from the $90\%$ and $98\%$ ranges plotted in Fig.~\ref{fig:AutoCorrResults_chi2}.
This information is also given in Table~\ref{tab:PSTab}. Calculating the PSD on the observed \textit{AMS-02} 
residual positron fraction gives us the black line, which is very close to the median noise mock simulation 
(blue dashed line). Thus, with the current data there is no indication in the \textit{AMS-02} data for a deviation 
from a smooth spectrum, just relying on the PSD $\chi^{2}$ criterion.

\begin{table}[t]
    \begin{tabular}{cccc}
         \hline
           Type of ranking & $\%$inc. f. & $\%$inc. f. & $\%$inc. f. \\
               & (16$\%$) & (5$\%$)  & (1$\%$)  \\
            \hline \hline
            $\chi^{2}/$d.o.f. on Power Spectrum  & 19.8 & 7.2 & 1.8 \\
            $\Sigma_{i=-30}^{i=+30} |h(f_{i})|^{2}$ & 42.0 & 23.7 & 7.6 \\
            1 - ($\Sigma_{i=-15}^{i=+15} |h(f_{i})|^{2}$/$\Sigma_{i=-30}^{i=+30} |h(f_{i})|^{2}$) & 22.8 & 9.7 & 2.9 \\
            \hline \hline 
        \end{tabular}
        \vspace{0.0cm}
\caption{The potential to observe power from small-scale features to the residual  \textit{AMS-02} positron 
fraction after 6.5 years of observations. We use the energy range between 5 and 500 GeV, with $f=1/ln(E/(1 \; \textrm{GeV}))$ in 
the range of $\pm 30$, i.e $\pm1/2$ the number of logarithmically spaced E-bins.  For the first two criteria used to rank 
our 5670 Milky Way pulsar realizations, we give the fraction in $\%$, of these realizations that fall inside the the upper 
16$\%$, 5$\%$ and 1$\%$ noise ranges (``$\%$inc. f.''). For the third criterion, we give the lower 16$\%$, 5$\%$ and 1$\%$ 
noise ranges.} 
\vspace{-0.7cm}
\label{tab:PSTab}
\end{table}
 
While the  $\chi^{2}/$d.o.f. on the PSD criterion separates some of the pulsar population simulations from the simulations of 
noise around 
a smooth spectrum, it still leaves a significant level of overlap between the two. As shown in Table~\ref{tab:PSTab}, for the 
overwhelming number of pulsars simulations their features would not give a $\chi^{2}/$d.o.f. fit much different to what noise 
would. For that reason we seek alternative criteria to break the two sets of simulations apart. 

\begin{figure*}
\begin{center}
\begin{tabular}{c c}
\hspace{-0.15in}
\includegraphics[width=3.55in,angle=0]{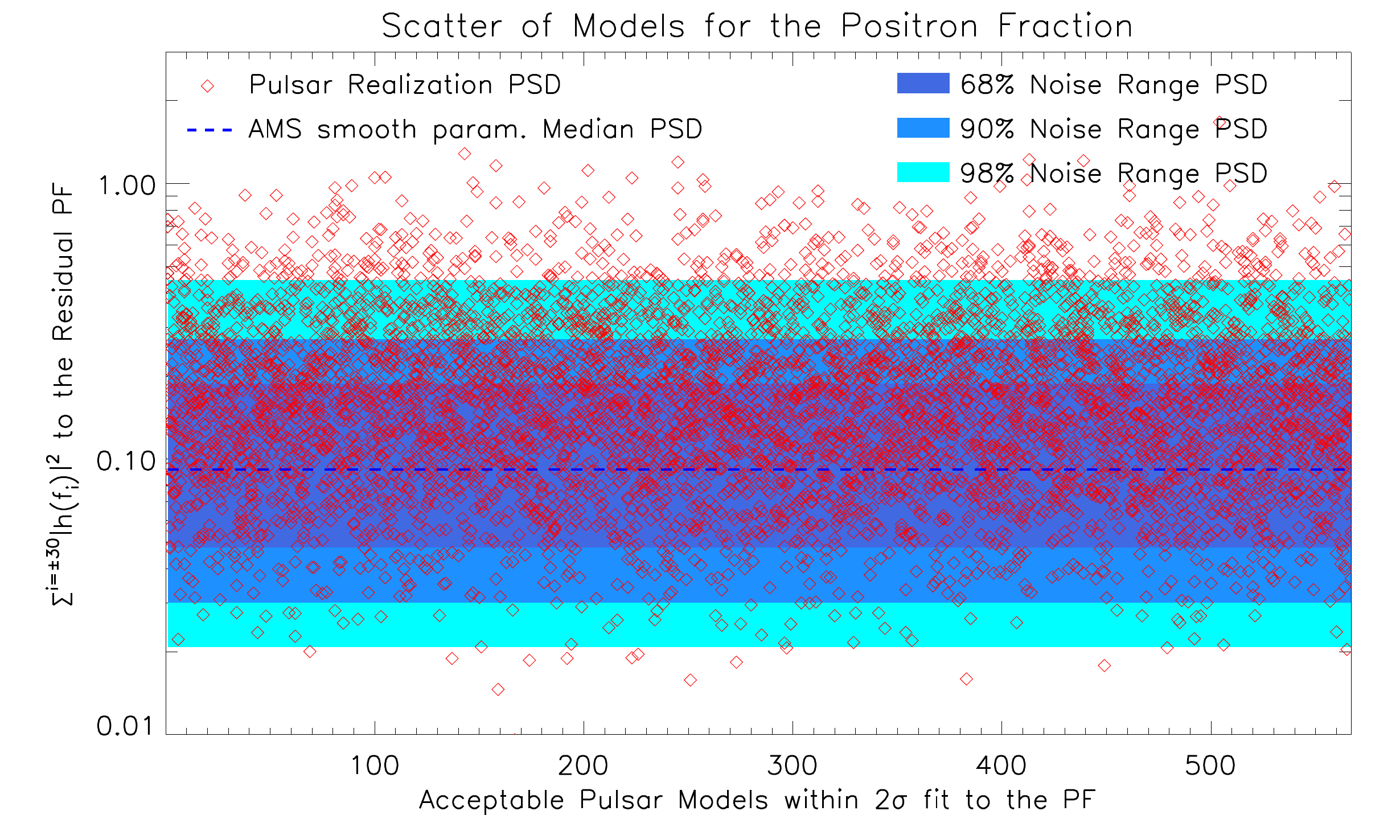}
\hspace{-0.0in}
\includegraphics[width=3.55in,angle=0]{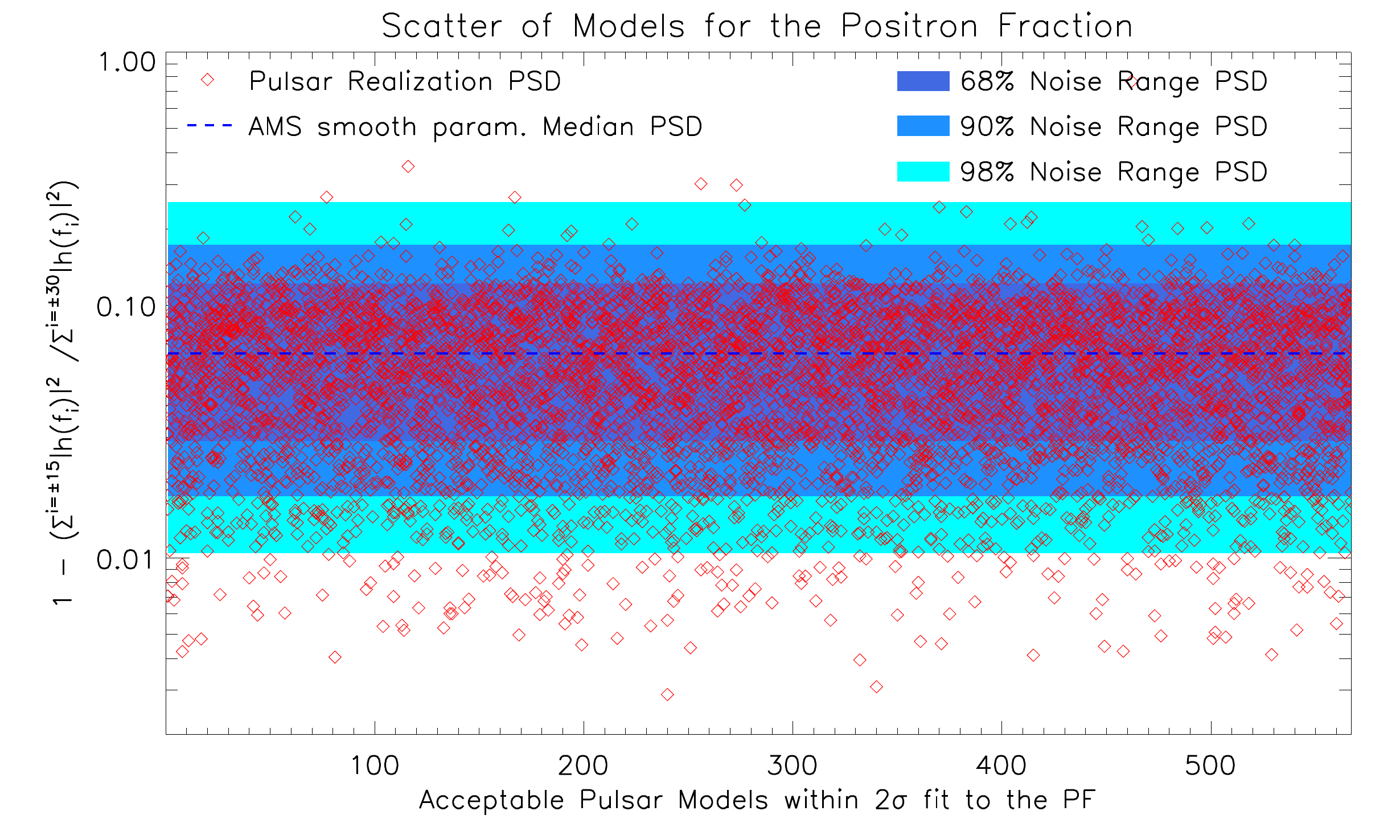}
\end{tabular}
\end{center}
\vspace{-0.55cm}
\caption{As with Fig.~\ref{fig:AutoCorrResults_chi2}, the scatter of the simulated Milky Way pulsars observational realizations, 
plotted by different characteristics of their PSDs. In red diamonds, we show 10 realizations from each of the 567 pulsar population simulations 
used. Left panel: each diamond gives the total PSD power $\Sigma_{i=-30}^{i=+30} |h(f_{i})|^{2}$, for each realization. The blue 
bands show the ranges of PSD total power for an underlying smooth spectrum with the addition of \textit{AMS-02} noise and the 
blue dashed line the median expectation. Spectral features from pulsars increase the total PSD power, as shown by the overall 
shift of the diamonds with respect to the blue bands.
Right panel: each red diamond gives the ratio of PSD power in the lower half modes to the total PSD power. To show how the 
pulsars predictions differ from that of noise, we plot the 1 - ($\Sigma_{i=-15}^{i=+15} |h(f_{i})|^{2}$/$\Sigma_{i=-30}^{i=+30} |h(f_{i})|^{2}$.
As with the left panel, the blue bands show the \textit{AMS-02} noise relevant ranges and the blue dashed line the median expectation. 
Pulsars predict a greater fraction of PSD power in the lower modes than regular noise. The diamonds are shifted to lower values in 
the $y$-axis with respect to the blue bands (see also text and Table~\ref{tab:PSTab}).}
\label{fig:AutoCorrResults}
\end{figure*}

In Fig.~\ref{fig:PSD_vs_frequency}, we showed two Milky Way pulsars simulations that compared to noise, give an increased 
total power in the power spectrum, i.e. a larger $\Sigma_{i=-30}^{i=+30} |h(f_{i})|^{2}$. In Fig.~\ref{fig:AutoCorrResults} 
(left panel), we plot the scatter of the 5670 realizations of pulsar population simulations in terms of that total power. Our $x$-axis 
is as with Fig.~\ref{fig:AutoCorrResults_chi2}, i.e. ranks the models from better to worse, in terms of their ability to fit the 
\textit{AMS-02} positron fraction measurement. There is no clear pattern on the total power in the PSD (our $y$-axis) a 
pulsar population simulation gives 
versus the quality of fit it has on the \textit{AMS-02} data. All these simulations provide a relatively good quality of fit to the 
\textit{AMS-02} positron fraction spectrum, as they give up to a $\chi^{2}/$d.o.f.=1.29. Simulations that would predict very large
 spectral features are excluded. There is a large scatter along the $y$-axis, $\Sigma_{i=-30}^{i=+30} |h(f_{i})|^{2}$ values. This
is directly related to the relatively large noise present in that energy range. We tried an alternative, narrower energy range and 
concluded that using the 5-500 GeV data is close to the optimal choice in searching for signals of spectral features given, the 
span of possible pulsar models explaining the data. Even with the large scatter along the $y$-axis, our pulsar population 
simulations give a larger total power than the regular noise around a smooth positron fraction spectrum does. In the blue 
shaded bands of the left panel of Fig.~\ref{fig:AutoCorrResults_chi2}, we give the $68\%$, $90\%$ and $98\%$ two-sided 
ranges on the $\Sigma_{i=-30}^{i=+30} |h(f_{i})|^{2}$ from the 1000 noise simulations. We find that of the 5670 observation 
realizations, 7.6$\%$ and 23.7$\%$ lie outside the $99 \%$ and $95 \%$ upper band end along the respective $y$-axis. Using 
the total power is quite a more sensitive criterion in separating the pulsar population simulations from noise. We note that 
still a significant fraction of pulsar population simulations, predict too little additional structure in the 5-500 GeV range of the 
positron fraction, to give a signal in the PSD.  
  
In the right panel of Fig.~\ref{fig:AutoCorrResults_chi2}, we explore further the point that pulsar realizations have a larger fraction 
of their power in low frequency $f=1/ln(E/(1 \; \textrm{GeV}))$ modes, compared to noise simulations. That is directly related to 
the fact that pulsars predict spectral features that span only a small number or energy bins. Our $x$-axis is the same as with the
left panel on the same figure (unique pulsar population simulations that fit the positron fraction). Our $y$-axis, gives the 1 - 
($\Sigma_{i=-15}^{i=+15} |h(f_{i})|^{2}$/$\Sigma_{i=-30}^{i=+30} |h(f_{i})|^{2}$). Pulsars with a larger fraction of their power
in the low modes are at the bottom of the $y$-axis. The blue bands give the two sided $68\%$, $90\%$ and $98\%$ ranges on the 
1 - ($\Sigma_{i=-15}^{i=+15} |h(f_{i})|^{2}$/$\Sigma_{i=-30}^{i=+30} |h(f_{i})|^{2}$) fraction evaluated from the 1000 noise simulations. 
Pulsar realizations are indeed shifted to lower values of that fraction. We find that 2.9$\%$ and 9.7$\%$ of the 5670 pulsars realizations
lie outside the $99 \%$ and $95 \%$ lower band end along the $y$-axis (see also Table~\ref{tab:PSTab}). Finally,  we note that there 
is no clear pattern in the goodness of fit that a pulsar population simulation has on the positron fraction spectrum and our $y$-axis value.

\subsection{Cross-correlation Analysis on Pulsar Population Simulations}
\label{subsec:Result_CC}

As we showed in Fig.~\ref{fig:ElecPos_Residuals_PulsarSim} for pulsar model I, there is significant noise in the residual electron 
and positron spectra of our pulsar population simulations after 6.5 years of \textit{AMS-02} observations. This results in the 
correlation coefficient 
$r_{xy}(m)$ to randomly acquire large positive and negative values for large values of shift $m$, correlating energy bins 
between the two residual spectra that are far away from each other. If pulsars are the underlying source for the high energy positrons, 
there is only one true realization of them. Since we don't know that, we want to search for general patterns among the many pulsar 
population simulations that are still in agreement with the cosmic-ray observations. 

\begin{figure}
\begin{center}
\vspace{-0.08in}
\hspace{-0.15in}
\includegraphics[width=3.55in,angle=0]{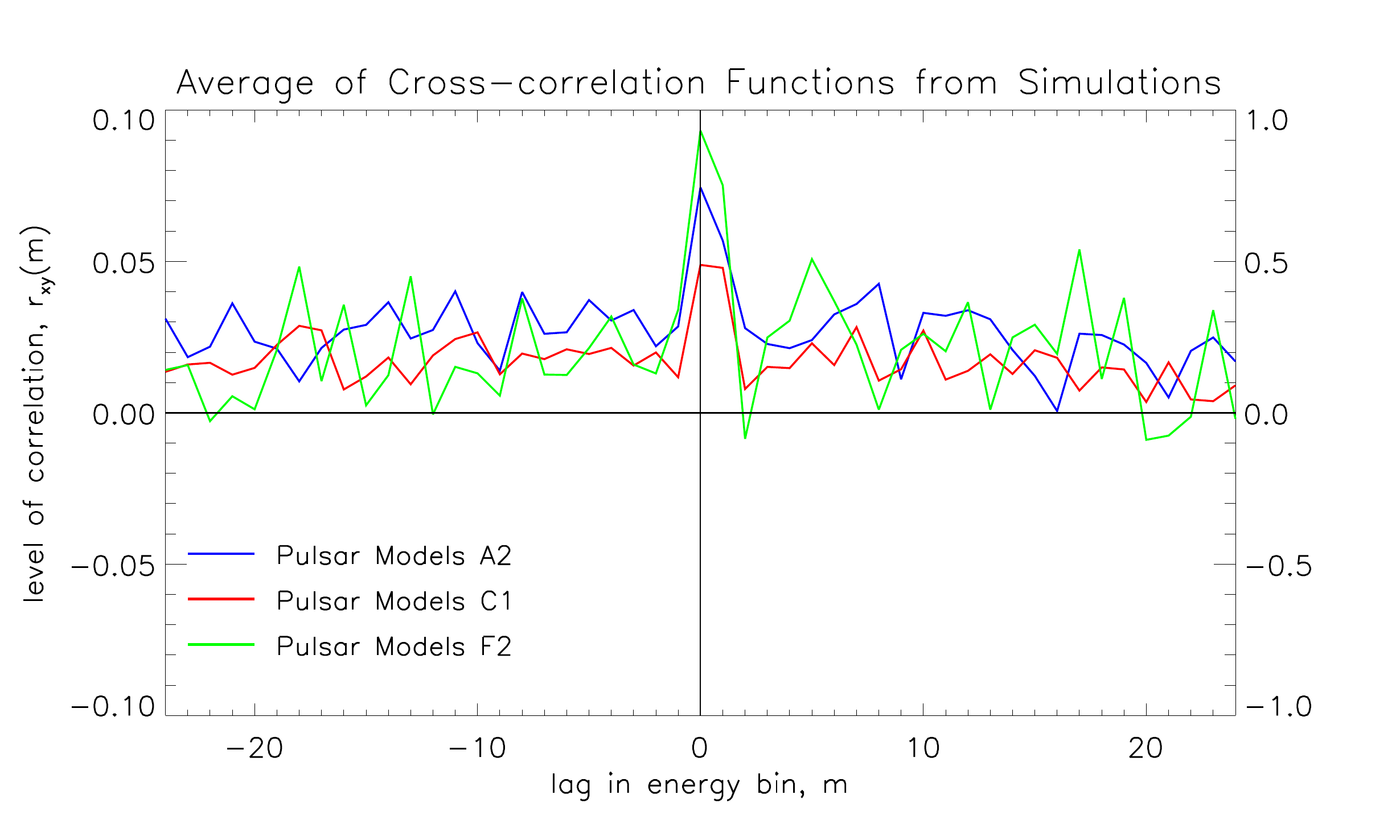}
\end{center}
\vspace{-0.20in}
\caption{The \textit{averaged} cross-correlation function between the electron and positron residual fluxes, evaluated from the observational realizations of many pulsar population simulations. We show the averages from the simulations built with ISM 
assumptions ``A2'', ``C1'' and ``F2'' (see text for details).}
\label{fig:CC_Pulsars_simulations}
\end{figure}

In Fig.~\ref{fig:CC_Pulsars_simulations}, for specific assumptions on the local ISM propagation conditions, we calculate the 
average of the cross-correlation functions of all realizations produced for pulsar population simulations under the same 
ISM assumptions. In Ref.~\cite{Cholis:2021kqk}, the pulsar population simulations tested to the cosmic-ray observations, 
were created for 12 different combinations
of choices for the local cosmic-ray diffusion and averaged energy losses (see Table II of Ref.~\cite{Cholis:2021kqk}). The 567 
pulsar population simulations in agreement with the cosmic-ray spectral data with their 5670 observational realizations can 
thus be partitioned 
in 12 groups based on these ISM assumptions. As was shown in Ref.~\cite{Cholis:2021kqk}, the combination of diffusion and 
energy losses assumptions can have an important impact on the quality of fit pulsar population simulations give to the 
cosmic-ray electron 
and positron observations. It can also have an effect on the presence or absence of prominent spectral features, of interest 
in this study. 
In Fig.~\ref{fig:CC_Pulsars_simulations}, the averaged cross-correlation function between the residual electron and positron  
fluxes is shown for three cases of ISM assumptions. In blue, we give give the averaged $r_{xy}(m)$ for the 440 realizations of 
pulsar population simulations produced under the ``A2'' ISM assumption. In the red line we give the equivalent averaged $r_{xy}(m)$ for 
the ``C1'' ISM assumption coming from 920 realizations and in green we give the averaged $r_{xy}(m)$ for the ``F2'' ISM assumption 
coming from 140 realizations. As can be seen for all averaged cross-correlation functions, they peak at $m=0$ with $m=+1$,
being the second highest point.

Models ``A'', take a local diffusion scale hight of $z_{L} = 5.7$ away from the galactic disk. As we described in Section 
\ref{subsec:Pulsar_Simulations}, the diffusion is assumed to be isotropic and homogeneous and given by a rigidity-dependent 
diffusion coefficient. Models ``A'', take the diffusion coefficient $D_{0} = 1.40\times10^{2} \, \textrm{pc}^2$/kyr  at 1 GV and 
the diffusion index $\delta = 0.33$. Models ``C'', 
have instead $z_{L} = 5.5$, $D_{0} = 0.921\times10^{2} \, \textrm{pc}^2$/kyr and $\delta = 0.40$, i.e. slower diffusion for lower 
rigidity cosmic rays but also becoming faster in a more rapid manner with increasing rigidity. Finally, models ``F'' take $z_{L} = 3.0$, 
$D_{0} = 0.337\times10^{2} \, \textrm{pc}^2$/kyr and $\delta = 0.43$, i.e. assume that cosmic rays once reaching only a 3 kpc 
distance away from the disk they will escape. Also, models ``1'' take for the high energy cosmic rays their energy loss's rate 
proportionality coefficient $b$ to be $b=5.05 \times 10^{-6} \textrm{GeV}^{-1} \textrm{kyr}^{-1}$. This represents conventional
assumptions for the local interstellar medium radiation field and magnetic field \cite{galprop, Moskalenko:2005ng, 
Porter:2005qx, Porter:2017vaa, 2012ApJ...761L..11J}. Instead, models ``2'' take 
a higher energy loss rate with $b=8.02 \times 10^{-6} \textrm{GeV}^{-1} \textrm{kyr}^{-1}$, that is still within the estimated 
relevant uncertainties. 

\begin{figure}
\begin{center}
\vspace{-0.08in}
\hspace{-0.15in}
\includegraphics[width=3.55in,angle=0]{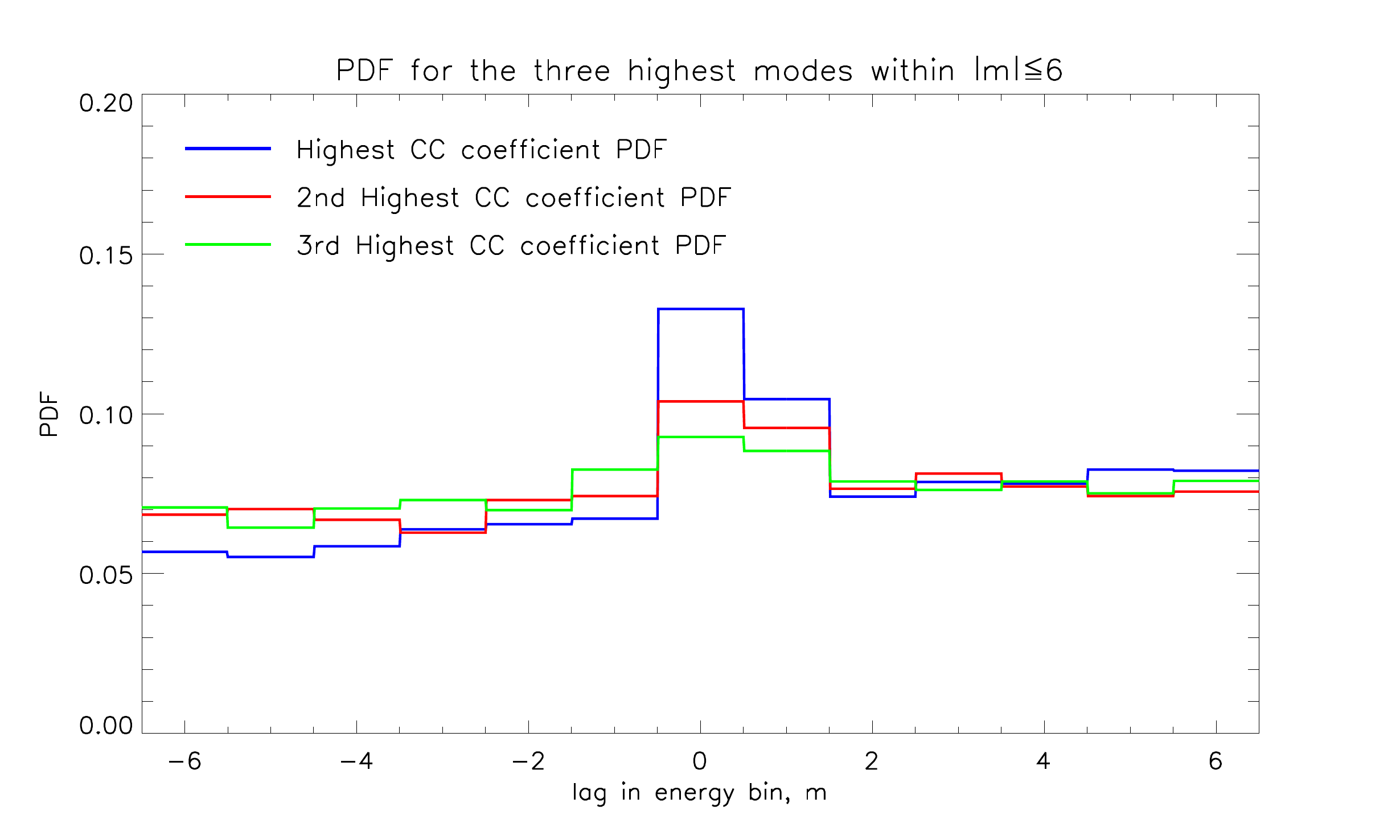}
\end{center}
\vspace{-0.20in}
\caption{The distribution of the shift $m$ of the highest (blue), 2nd highest (red) and 3rd highest (green) cross-correlation 
coefficient evaluated from our simulations, as those presented in Fig.~\ref{fig:ElecPos_Residuals_PulsarSim} (bottom panel). 
We rank only the range within $|m| \leq 6$, as higher values of $|m|$ have large noise. Each of the histogram distributions is 
normalized to give a total area of 1, thus these are PDFs. We use the entire sample of 5670 pulsar realizations in evaluating 
these PDFs.}
\label{fig:CC_Pulsars_PDF}
\end{figure}

In Fig.~\ref{fig:CC_Pulsars_PDF}, we show for the entire set of 5670 realizations, the distribution of the location of the shift $m$, 
for which the three highest values of the cross-correlation coefficient $r_{xy}$ appear. We study only the region within $|m| \leq 6$, 
to avoid the impact of correlating the noisy data at higher energies with features at significantly lower energies. Each of our three 
histograms is normalized to have an area of 1, thus making these probability density functions (PDFs) for the occurrence of the 
highest (blue line), second highest (red line) and third highest (green line) cross-correlation coefficient value. Again, the highest 
value for $r_{xy}(m)$ is for $m=0$, with the second most likely shift being for $m=+1$, i.e. the pulsar features appear on the 
electrons to be coinciding in energy or shifted by one bin at lower energies compared to the positrons. A similar effect is seen for the second and third highest
values of $r_{xy}$, but in a less prominent manner. The distribution of less high values of the $r_{xy}$ is fairly flat with $m$. As the 
\textit{AMS-02} collects more data and the noise gets decreased, the correlations of underlying features (if those are present)
will become more easy to identify. 

\begin{figure}
\vspace{-0.10in}
\hspace{-0.15in}
\includegraphics[width=3.72in,angle=0]{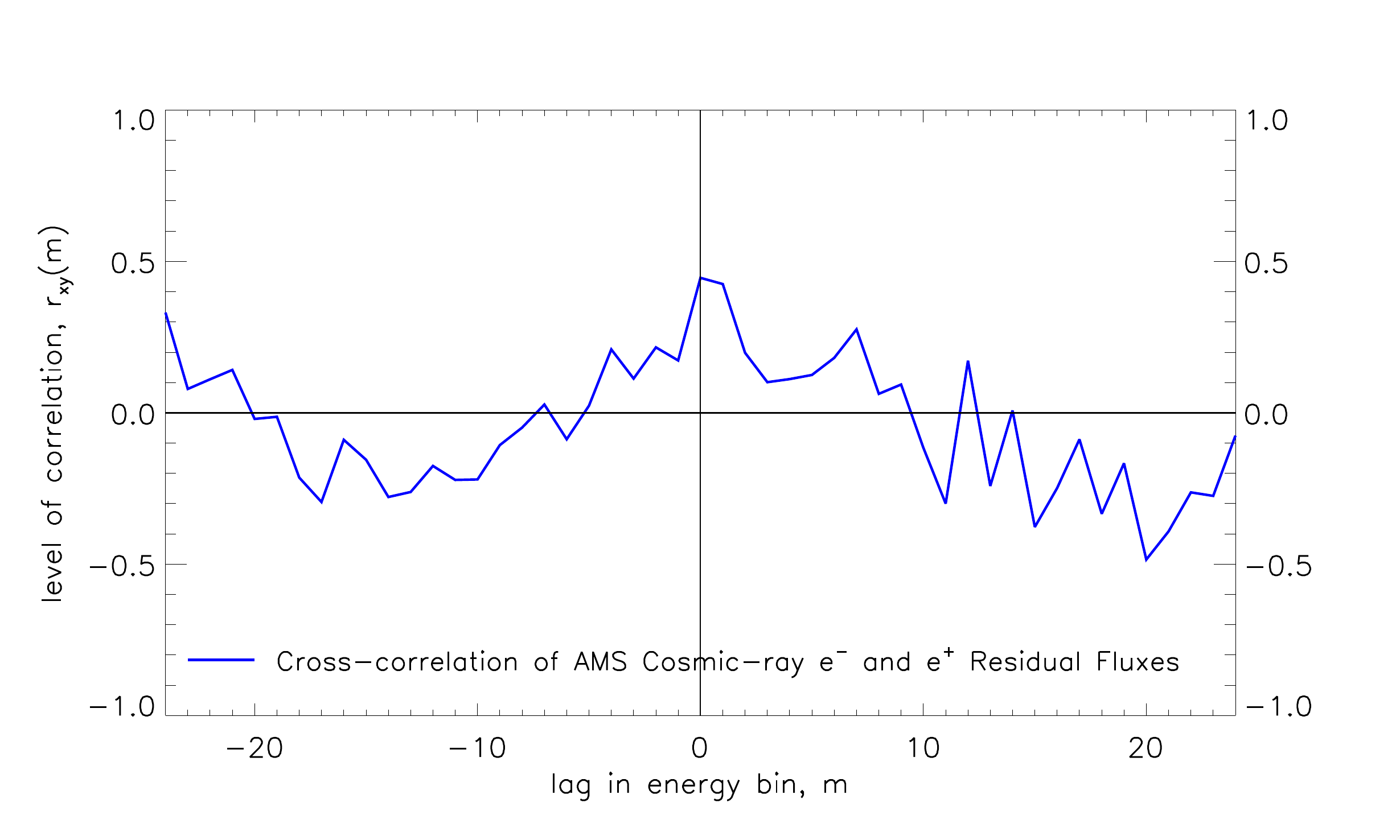}
\vspace{-0.28in}
\caption{The cross-correlation function between the \textit{AMS-02} electron and positron residual fluxes. The vertical, guide line is 
at $m = 0$. The correlation peaks at $m=0$ similarly to our pulsars expectation.}
\label{fig:CC_AMS_data}
\end{figure}

In Fig.~\ref{fig:CC_AMS_data}, we show the cross-correlation analysis of the \textit{AMS-02} observations of Refs.~\cite{AMS:2019iwo,
 AMS:2019rhg}, using the cosmic-ray electron and positron fluxes that we have shown in Fig.~\ref{fig:ElecPos_Residuals} and 
discussed in section~\ref{subsec:CC_on_EP}. Interestingly, we find a positive correlation between the residual spectra at $m=0$, 
with the second higher local value being at $m=+1$ as expected by our pulsar population simulations. Moreover, at the 
higher values of $|m|$, we do find large values of $r_{xy}$ that we associate to cross-correlating features at high energies from one
residual spectrum to lower energies of the other residual spectrum.  

We consider the result of Fig.~\ref{fig:CC_AMS_data}, in association with all our expectations from the pulsar populations as we 
described them in this section and in section~\ref{subsec:CC_on_EP} to be an intriguing finding. We may be at the point of 
detecting signs of correlated in energy spectral features in the \textit{AMS-02} data. If that is the case, then with longer observations,
we will see that correlation signal become more robust and we may also see a signal of those features in the power spectrum of
the residual positron fraction as discussed in sections~\ref{subsec:PS_on_PF} and~\ref{subsec:Result_PS_on_PF}.  

\section{Conclusions and Discussion}
\label{sec:Conclusions}

In this paper, we adapt and implement a power-spectrum and a cross-correlation analysis, on the cosmic-ray measurements 
form the \textit{AMS-02}, on the positron fraction and the electron and positron fluxes. We search for signals of underlying 
spectral features in these measurements. Such spectral features can exist if a population of relatively local Milky Way pulsars 
(or SNRs) is to explain the rising cosmic-ray positron fraction and the respective hardening of the cosmic-ray positron flux 
spectrum. Powerful and young to middle-aged pulsars, can be significant cosmic-ray sources on the positron fraction and 
electron and positron flux spectra, giving features localized in energy as we show in Figs.~\ref{fig:PositronFraction} 
and~\ref{fig:Electrons_and_positrons}.

We implement a power-spectrum analysis on the cosmic-ray positron fraction, focusing on the capacity the current 
measurements have in finding a signal of spectral features from pulsars. To avoid any bias on our results from the large scale 
evolution of the positron fraction spectrum, we subtract the smoothed positron fraction spectrum and evaluate the power-spectral 
density on its residual spectrum, that would still retain the smaller scale features we are after. We do that using a vast library 
of Milky Way pulsar population models from 
Ref.~\cite{Cholis:2021kqk}. As we describe in more detail in section~\ref{subsec:Pulsar_Simulations}, those simulations account 
for all the relevant astrophysical modeling uncertainties relating to the pulsar properties as cosmic-ray electron and positron sources, 
the stochastic nature of their birth, their initial power and their surrounding environment, affecting the cosmic-ray spectra they 
inject to the ISM. Those simulations also account for uncertainties on the characteristics of the other components of the cosmic-ray 
electrons and positrons, namely the primary and secondary fluxes, and finally they model the uncertainties of cosmic-ray propagation 
through the ISM and the heliosphere. The models that we use have already been tested in Ref.~\cite{Cholis:2021kqk} for their 
compatibility to the electron and positron measurements made by \textit{AMS-02}, \textit{CALET} and \textit{DAMPE}. We use 
only a subset of 567 simulations that give good fits to the observations from the original $\simeq 7.3 \times 10^{3}$ simulations 
produced in~\cite{Cholis:2021kqk}. Moreover, to account for the presence of statistical noise in the \textit{AMS-02} positron fraction 
measurement, for each of the 567 pulsar population simulations we generate 10 realizations. Thus we test our hypothesis on 
performing a power-spectrum analysis to search for spectral features from pulsars on 5670 realizations of the positron fraction 
spectrum, as that would be measured after 6.5 years of observations, in agreement with the relevant published results 
of~\cite{AMS:2019iwo}.

As we show in Fig.~\ref{fig:PSD_vs_frequency} and discuss in section~\ref{subsec:PS_on_PF}, by using the residual of the 
positron fraction between 5 and 500 GeV and calculating PSDs, there are specific ``frequency'' $f = 1/ln(E/\textrm{GeV})$ modes,
where a signal of the presence of the pulsar induced spectral features would appear. Any power-spectrum analysis would only 
find signals of underlying features in certain modes, but would not retain the information on the actual energy they occur. To 
understand the ability of a power-spectrum analysis to find such signals, we compare the PSDs of the 5670 pulsar population 
realizations to the PSDs expected by just having noise on an otherwise smooth spectrum for which we can evaluate its residual 
(fluctuating around zero but with larger noise at high energies). We create 1000 such simulated AMS-02 positron fraction 
measurements, assuming a smooth positron 
fraction measurement, which we then subtract and after evaluate 1000 PSDs, giving us the PSD-ranges due to the statistical noise.
As we show in our results section~\ref{subsec:Result_PS_on_PF} and in particular in Fig.~\ref{fig:ElecPos_Residuals_PulsarSim} 
and Table~\ref{tab:PSTab}, the best way to search for pulsar features through a power-spectrum analysis, is to evaluate the total 
PSD power on the calculated residual positron fraction. A significant fraction of pulsar population simulations predict an increased 
total PSD power compared to simple noise. Also, the pulsar population simulations predict that the ratio of PSD power in the lower 
half modes to the total PSD power is increased as well compared to just noise simulations. 

Using the same 567 pulsar population simulations as in our power-spectrum analysis, we  also perform a cross-correlation analysis.
From each simulation, we evaluate the expected cosmic-ray electron and positron fluxes after 6.5 years of \textit{AMS-02} 
observations, comparable to \cite{AMS:2019iwo, AMS:2019rhg} and produce 10 realizations for each of the electron and positron 
fluxes. We then evaluate the relevant smooth spectra and derive the residual electron and positron fluxes as shown in 
Fig.~\ref{fig:ElecPos_Residuals_PulsarSim}; which we then cross-correlate (discussed in section~\ref{subsec:CC_on_EP}).
We find that even with the noise present, some of the underlying common spectral features on the electron and positron fluxes 
predicted by pulsars will remain. That results in having a positive cross-correlation signal between the residual electron and 
residual positron fluxes. As we show in Figs.~\ref{fig:CC_Pulsars_simulations} and ~\ref{fig:CC_Pulsars_PDF}, for the entire 
set and for subsets of our simulations, a positive correlation signal can exist for at least 1/4th of our simulations. That signal 
would suggest that spectral features on the residual cosmic-ray electrons will coincide in energy to spectral features in 
positrons. Also, a positive correlation signal can exist suggesting that the electron spectral features on the residual flux, are 
shifted by one bin at lower energies compared to the spectral features at the positrons. The latter has to do with the electron 
spectrum in its non-residual version being softer than the positron one. 

Finally, we perform a cross-correlation analysis on the \textit{AMS-02} electron and positron fluxes. By first evaluating the 
relevant residual spectra that we discuss in detail in section~\ref{subsec:CC_on_EP}, and show in Fig.~\ref{fig:ElecPos_Residuals}, 
we then implement the same cross-correlation technique as we did for our pulsar population simulations. In Fig.~\ref{fig:CC_AMS_data},
we show that we find a positive correlation between the \textit{AMS-02} residual electron and positron spectra. Similar to our 
expectations from our pulsar population simulations, in the \textit{AMS-02} data, there is clear indication for a positive 
correlation between 
these spectra that suggests their underlying spectral features coincide in energy. Furthermore, there is a slightly less 
prominent positive correlation between the residual positron flux and the residual electron flux shifted by one bin at lower 
energies (again as expected in pulsars). We find these results intriguing as we may be at the verge of observing signals of
spectral features in the cosmic-ray electrons and positrons, as has been hypothesized for decades now (see e.g. 
\cite{1995A&A...294L..41A, Malyshev:2009tw, Profumo:2008ms, Kawanaka:2009dk, Grasso:2009ma, Cholis:2017ccs, 
Orusa:2021tts}). This would also be a significant indicator against the dark matter interpretation of the rising positron 
fraction. 

In the future we expect even higher quality measurements, due to increased data-taking periods by \textit{AMS-02} and
even better-controlled systematics. Also, future developments as the hypothetical AMS-100 detector \cite{Schael:2019lvx}, can 
transform the quality of the measured cosmic-ray fluxes. Both the power-spectrum analysis on the residual positron 
fraction and the cross-correlation between the residual electron and positron fluxes may 
provide robust evidence of underlying spectral features from powerful local cosmic-ray 
sources.

 \textit{Acknowledgements:} 
We would like to thank Sawyer Hall and Tanvi Karwal for valuable discussions at the early stages of this work. We also thank 
Marc Kamionkowski, Ely Kovetz and Iason Krommydas for useful discussions. We also acknowledge the use of \texttt{GALPROP} \cite{galprop}.
IC acknowledges support from the Michigan Space Grant Consortium, NASA Grant No. 80NSSC20M0124.
IC acknowledges that this material is based upon work supported by the U.S. Department of Energy, Office of Science, Office of High Energy Physics, under Award No. DE-SC0022352.

\bibliography{CR_AC_and_CC}

\end{document}